%% file: my.tex
\documentclass{elsart}

\newcommand{\mysection}{\setcounter{equation}{0}\section}
\renewcommand{\theequation}{\thesection.\arabic{equation}}
\begin{document}

\include{my0}  
\include{my1}  
\include{my2}  
\include{my3}  
\include{my4}  
\include{my5}  
\include{my6}  
\include{my7}  
\include{my8}  
\include{my9}  

\appendix
\include{myaa} 
\include{mybb} 

\include{myr}  
\include{myf}  
\end{document}

%% file: my0.tex
\vskip 0.2cm
\hfill{YITP-SB-01-10}
\vskip 0.2cm
\centerline{\large\bf {Evolution program for parton densities with}}
\centerline{\large\bf {perturbative heavy flavor boundary conditions}}
\vskip 0.2cm
\centerline {A. Chuvakin, J. Smith}
\centerline{\it C.N. Yang Institute for Theoretical Physics,}
\centerline{\it State University of New York at Stony Brook,
New York 11794-3840.}
\vskip 0.2cm
\centerline{February 2001}
\vskip 0.2cm
\centerline{\bf Abstract}
\vskip 0.3cm

A new code for the scale evolution of 
modified-minimal-subtraction-scheme parton densities is described. 
Through next-to-leading order the program uses exact 
splitting functions. In next-to-next-to-leading order 
approximate splitting functions are used. 
For efficiency the program includes analytical results for the evaluation 
of the weights required for the integrations over the longitudinal
momentum fractions of the partons.  It also incorporates the 
operator matrix elements required for the matching conditions 
across heavy flavor thresholds in higher order perturbation theory. 
The more efficient handling of the weights implies that the code 
is faster than similar evolution codes in all
modes of operation. The program is written in the C programming language.



%% file: my1.tex

\tableofcontents

\mysection{Program Summary}

{\it Title of Program:} ADENS \\
\noindent
{\it Computer: }  AlphaStation 4/500 \\
\noindent
{\it Operating system:} OSF V3.2 \\
\noindent
{\it Programming language:} C \\
\noindent
{\it Number of lines in distributed program:} 11658 
\noindent
{\it Nature of physical problem:}
\noindent
solution of the parton density evolution equations with LO, NLO and
NNLO splitting functions and NLO, NNLO 
heavy flavor threshold matching conditions\\
\noindent
{\it Method of solution:}
\noindent
$x$-space integration with analytic evaluation of weights\\
\noindent
{\it Typical running time:} see table in Section 7 \\
\noindent
{\it The program www site:} \\
http://insti.physics.sunysb.edu/$\sim$chuvakin/adens-24.0.tar.gz \\
http://insti.physics.sunysb.edu/$\sim$smith/adens-24.0.tar.gz \\


%% file: my2.tex

\tableofcontents
\mysection{Introduction}


Deep-inelastic lepton-hadron scattering experiments 
probe the internal structure of hadrons.  
The lepton-hadron inclusive cross sections
may be written in terms of structure functions, which 
depend on the virtuality of the probe $Q^2$.
Three structure functions $F_1$, $F_2$ and $F_L$
are necessary to describe neutral current (photon and $Z$-boson
exchange) and charged current ($W$-boson exchange) reactions. 
In perturbative quantum chromodynamics (pQCD) 
the probe interacts with partonic constituents of the hadron. 
There are probability densities $f(x,\mu^2)$ 
to find partons carrying a fraction $x$ ($0 < x \le 1$) of the 
longitudinal momentum of the hadron at a mass factorization scale $\mu$.  
Therefore the $F_i$, $i=1,2,L$ also depend on $x$ and $\mu$.

The operator product expansion (OPE)
allows the structure functions to be written as convolutions of 
the parton (quark and gluon) probability densities 
with partonic hard scattering cross sections 
(or coefficient functions). The latter can be calculated in pQCD.
Even though the former cannot be calculated in pQCD, 
their $\mu$ dependence is determined by a set of 
integro-differential equations, 
the (Dokshitzer-Gribov-Lipatov)-Altarelli-Parisi
evolution equations \cite{ap}, which follow from 
renormalization group analysis. 
Discussions of the pQCD description of deep inelastic 
scattering reactions are available in \cite{rob} and \cite{esw}.
The probability densities and splitting functions are defined
in the modified-minimal-subtraction ($\overline{\rm MS}$) scheme.

For simplicity we will call the above equations the evolution equations. 
They describe processes where a massless parton 
(quark or gluon) carrying a fraction of the longitudinal 
momentum of the incoming hadron radiates
a massless parton and becomes a (different) 
massless parton with a different momentum fraction.
The probability for this process to happen is determined 
by splitting functions which are computed order-by-order
in pQCD. The leading-order (LO) and next-to-leading order (NLO)
 splitting functions have been known for some time \cite{gp}, \cite{gw},
\cite{frs}, \cite{gry}, \cite{fkl}, \cite{hn} and the results
are summarized in a convenient form in \cite{esw}.
Recently some moments of the next-to-next-to-leading order
(NNLO) splitting functions have been 
calculated in \cite{mom}, see also \cite{grac} and \cite{nevo} .
If the $x$-dependence of the quark and gluon densities 
in a hadron are parametrized at one value of $\mu$, (say at $\mu_0$,)
then the solutions of the evolution equations with the
above LO, NLO or NNLO splitting functions yield the 
$x$ dependence of the massless parton densities at a different $\mu$.
There is a second scale in the pQCD theory, 
the renormalization scale, which appears in argument of the
the running coupling $\alpha_{s}$. 
It is usually set to be the same as the mass factorization scale $\mu$
so $\alpha_s = \alpha_s(\mu^2)$. 

The flavor dependence of the quark and anti-quark densities is governed by 
the flavor group, which is SU(2) for the up and down quarks, SU(3) for the
up, down and strange quarks etc. Therefore is is convenient to
form flavor non-singlet and flavor (pure) singlet combinations
of densities. The
former have their own evolution equations. The latter mix with the gluons
and the combined evolution is described by matrices which obey 
coupled integro-differential equations.

A number of methods to solve the evolution equations for
the parton densities have been proposed, including 
direct $x$-space methods, \cite{miya}, \cite{botje},
\cite{pasc}, \cite{brnv}, \cite{mrst98}, \cite{cteq5},
orthogonal polynomial methods 
\cite {cfr}, \cite{cori}, and Mellin-transform methods \cite{riemer}, 
\cite{grv98}.
A compilation of parton density sets is available in
\cite{pb}. 

The best method, which should be both accurate and fast,
depends on region chosen in $x$ and $\mu^2$.
Currently the requirements are that the code be able to evolve 
densities from a minimum $\mu^2$ near 0.26 ${\rm GeV}^2$ up to
a maximum $\mu^2$ near $10^6$ ${\rm GeV}^2$ 
required for QCD studies for the future Large Hadron Collider at CERN.
The range in $x$ is from a minimum value 
near $10^{-5}$ up to a maximum near unity. We use 
the direct $x$-space method, with the following additional features. 

One of our aims is a better treatment of parton density evolution
for "light" $u$, $d$ and $s$ quarks near the heavy flavor 
thresholds chosen to be at the charm and the bottom 
quark masses ($m_c$ and $m_b$ respectively). The 
parton density description must be modified to incorporate
new $c$ and $b$ "heavy" quark densities as the evolution
scale increases. The implementation of the NLO and NNLO matching 
conditions across heavy flavor thresholds in the variable flavor 
number schemes (VFNS) \cite{acot}, \cite{csn}, \cite{cbot}, \cite{csh}  
involve large cancellations between
various terms in the expressions for the structure functions.
Poor numerical accuracy in the solution for the
evolution of the parton densities at small scales would 
spoil these cancellations and ruin the VFNS predictions.
We achieve the required accuracy by avoiding one numerical integration
in our program so we analytically calculate 
the weights for the exact LO, the exact NLO and 
the approximate NNLO splitting functions.
The approximate NNLO splitting functions are taken from  
\cite{3loop1},\cite{3loop2}, while the relevant operator matrix
elements (OMEs), which provide the matching conditions
on the parton densities across heavy flavor thresholds,
are taken from \cite{bmsn1}.

Since we start the scale evolution from a set of densities 
(input boundary conditions) at a low scale $\mu = \mu_0 \ll m_c$, 
the running coupling $\alpha_s(\mu^2)$ is large.  
We therefore use the exact solution of the NLO equation for 
$\alpha_s$ and match the values on both sides of the heavy 
flavor thresholds to three decimal places. 
We mention here that the NNLO matching conditions on $\alpha_s$ 
across heavy flavor thresholds are available in \cite{2loop} and \cite{2loop2}.
Our program evolves both light and heavy parton densities in LO, NLO 
and NNLO from a minimum $x$ equal to $10^{-7}$ to a maximum $x$
equal to unity, a mimimum $\mu^2 = 0.26$ 
$ ({\rm GeV})^2$ in LO and $\mu^2 = 0.40$ 
$ ({\rm GeV})^2$ in NLO and NNLO and and a maximum $\mu^2 = 10^6$ 
$ ({\rm GeV})^2$.
Results have been published in \cite{csn}, \cite{cbot}, 
\cite{csh} and \cite{cs}.  Here we give a detailed write up of the program.

%% file: my3.tex
%
\mysection{The evolution equations}
\subsection{Definitions of densities}

We evolve combinations of up ($u$), down ($d$), strange $(s)$, 
charm $(c)$ and bottom ($b$) quark densities
which transform appropriately under the flavor group. Hence we define
flavor-non-singlet valence quark densities by 
\begin{equation}
\label{eqn3.1}
f_{k-\bar k}(n_f,x,\mu^2) \equiv f_k(n_f,x,\mu^2) - f_{\bar k}(n_f,x,\mu^2) 
\, , \, k=u,d \,.
\end{equation}
The flavor-singlet quark densities 
\begin{equation}
\label{eqn3.2}
 f_q^{\rm S}(n_f,x,\mu^2) = \sum_{k=1}^{n_f} f_{k+\bar k}(n_f,x,\mu^2) 
\end{equation}
are defined in terms of the expression 
\begin{equation}
\label{eqn3.3}
f_{k+\bar k}(n_f,x,\mu^2) \equiv f_k(n_f,x,\mu^2) + f_{\bar k}(n_f,x,\mu^2) 
\, ,\,k=u,d,s,c,b\,,
\end{equation}
when $n_f = 5$. Then  the flavor-non-singlet sea quark densities are
\begin{equation}
\label{eqn3.4}
 f_q^{\rm NS}(n_f,x,\mu^2) = f_{k+\bar k}(n_f,x,\mu^2) 
-\frac{1}{n_f} f_q^{\rm S}(n_f,x,\mu^2) \,.
\end{equation}
These equations will be discussed further in the next section.

\subsection{The evolution equations}

A typical evolution equation is that for
a flavor-non-singlet parton density $f^{NS}(x,\mu^2)$
\begin{equation}
\label{eqn3.5}
{\partial \over {\partial\ln \mu^2}} \ f^{{NS}}(y,\mu^2)\ =
{{\alpha_s (\mu^2)} \over {2\pi}} \ \int_y^1{dx\over x} \
P_{{NS}}\bigl({y\over x}, \mu^2\bigr) \
f^{{NS}}(x,\mu^2) \ \ \ ,
\end{equation}
where $P^{{NS}}(y/x,\mu^2)$ is a non-singlet splitting function, and
$\alpha_s(\mu^2)$ is the running coupling.

The splitting functions in the evolution equations
can be expanded in a perturbation series 
in $\alpha_s$ into LO, NLO and NNLO terms as follows
\begin{equation}
\label{eqn3.6}
P(z,\mu^2)=  P^{(0)}(z,\mu^2)
+(\frac{\alpha_s(\mu^2)}{2\pi}) P^{(1)}(z,\mu^2)
+(\frac{\alpha_s(\mu^2)}{2\pi})^2 P^{(2)}(z,\mu^2).
\end{equation}
The non-singlet combinations of the $q_r(\bar q_r)$ to $q_s(\bar q_s)$
splitting functions, 
where the subscripts $r,s$ denote the flavors of the (anti)quarks 
$q$ and $\bar q$ respectively and
satisfy $r,s=1,\cdots,n_f$,
can be further decomposed into flavor diagonal parts proportional
to $\delta_{rs}$ and flavor independent parts.
In LO there is only one non-singlet splitting function $P_{qq}$
but in NLO it is convenient to form two combinations
from $P_{qq} $ and $P_{q \bar q}$ as follows
\begin{eqnarray}
\label{eqn3.7}
P_+=P_{qq}+P_{q{\bar q}}\,,
\nonumber \\
P_-=P_{qq}-P_{q{\bar q}}\,.
\end{eqnarray}
These splitting functions are used to evolve two independent types 
of non-singlet densities, which will be called plus and minus respectively. 
They are given by
\begin{eqnarray}
\label{eqn3.8}
f^+_i = f_q^{\rm NS}(n_f,x,\mu^2)\,,
\nonumber \\
f^-_j = f_{k-\bar k}(n_f,x,\mu^2)\,.
\end{eqnarray}
Since the general formulae in Eqs. (3.1)-(3.4) are rather involved
the easiest way to explain the indices is by explicitly
giving the combinations we use. For $j=1,2$ we have 
\begin{equation}
\label{eqn3.9}
f_1^-  =  u - \bar u\,, \, f_2^-  =  d - \bar d\,,
\end{equation}
which are used for all flavor density sets.
Then for three-flavor densities $i = 1,2,3$ and we define
\begin{eqnarray}
\label{eqn3.10}
&& f_1^+  =  u + \bar u - \Sigma(3)/3\,,\quad  \quad
   f_2^+  =  d + \bar d - \Sigma(3)/3\,, \nonumber \\
&& f_3^+  =  s + \bar s - \Sigma(3)/3\,, 
\end{eqnarray}
where $\Sigma(3) = f_q^S(3) = u + \bar u + d + \bar d + s + \bar s$.
These densities should be used for scales $\mu < m_c$.
For four-flavor densities $i = 1,2,3,4$ and we define
\begin{eqnarray}
\label{eqn3.11}
&& f_1^+  =  u + \bar u - \Sigma(4)/4\,, \quad \quad
   f_2^+  =  d + \bar d - \Sigma(4)/4\,, \nonumber \\
&& f_3^+  =  s + \bar s - \Sigma(4)/4\,, \quad \quad 
   f_4^+  =  c + \bar c - \Sigma(4)/4\,, 
\end{eqnarray}
where $\Sigma(4) = f_q^S(4) = c + \bar c + \Sigma(3)$.
These should be used for scales in the region $m_c \le \mu < m_b$.
For five-flavor densites $i = 1,2,3,4,5$ and we define 
\begin{eqnarray}
\label{eqn3.12}
&& f_1^+  =  u + \bar u - \Sigma(5)/5\,, \quad \quad 
   f_2^+  =  d + \bar d - \Sigma(5)/5\,, \nonumber \\
&& f_3^+  =  s + \bar s - \Sigma(5)/5\,, \quad \quad 
   f_4^+  =  c + \bar c - \Sigma(5)/5\,, \nonumber \\
&& f_5^+  =  b + \bar b - \Sigma(5)/5\,, 
\end{eqnarray}
where $\Sigma(5) = f^{S}_q(5) = b + \bar b + \Sigma(4)$.
These should be used for scales $\mu \ge m_b$.

If we define $t = \ln(\mu^2/(1\, {\rm GeV}^2)$ then  
we need to solve the four evolution equations 
\begin{eqnarray}
\label{eqn3.13}
\frac{\partial f^{+}_{i}(y,t)}{\partial t}
&=&\frac{\alpha _{s}(t)}{2\pi }\int_y^1 
\frac{dx}{x}P_{+}\bigl(\frac{y}{x},t\bigr)f^{+}_{i}(x,t) \,,
\end{eqnarray}
%
\begin{eqnarray}
\label{eqn3.14}
\frac{\partial f^{-}_{j}(y,t)}{\partial t}
&=&\frac{\alpha _{s}(t)}{2\pi }\int_y^1 
\frac{dx}{x}P_{-}\bigl(\frac{y}{x},t\bigr)f^{-}_{j}(x,t) \,,
\end{eqnarray}
%
\begin{eqnarray}
\label{eqn3.15}
\frac{\partial f_{g}(y,t)}{\partial  t}
&=&\frac{\alpha _{s}(t)}{2\pi }\int_y^1
\frac{dx}{x} \left[ P_{gq}\bigl(\frac{y}{x},t\bigr)f^{S}_{q}(x,t)
+P_{gg}\bigl(\frac{y}{x},t\bigr)f^{S}_{g}(x,t)\right] ,
\end{eqnarray}
%
\begin{eqnarray}
\label{eqn3.16}
\frac{\partial f^{S}_{q}(y,t)}{\partial  t}
&=&\frac{\alpha _{s}(t)}{2\pi }\int_y^1 
\frac{dx}{x} \left[ P_{qq}\bigl(\frac{y}{x},t\bigr)f^{S}_{q}(x,t)
+P_{qg}\bigl(\frac{y}{x},t\bigr)f^{S}_{g}(x,t)\right] ,
\end{eqnarray}
where for $\mu < m_c$ we set $i=1,2,3$, $j=1,2$, 
$f_q^S = \Sigma(3)$ and the gluon is a three-flavor gluon.
When $m_c \le \mu < m_b$, we use    
$i=1,2,3,4$, $j=1,2$, $f_q^S = \Sigma(4)$ and the gluon is a four-flavor
gluon.  Finally when $\mu \ge m_b$, we set
$i=1,2,3,4,5$, $j=1,2$, $f_q^S = \Sigma(5)$ and the gluon is 
a five-flavor gluon. Note that since NNLO splitting functions are approximate
we provide the high and low estimate for each splitting functions labeled A and
B. For all calculations we use their average so that the error is minimized.


The densities satisfy the momentum conservation sum rule
which we write in terms of the $u, d, .. b$ (anti)-quark 
and gluon densities as
\begin{eqnarray}
\label{eqn3.17}
&& \int_0^1 \, dx \, x
 \left[u(x,\mu^2) + \bar{u}(x,\mu^2)
    +  d(x,\mu^2) + \bar{d}(x,\mu^2)  \right. 
\nonumber \\ &&
\left.+ s(x,\mu^2) + \bar s(x,\mu^2)    
      + [c(x,\mu^2) + \bar c(x,\mu^2)] \theta(\mu^2 - m_c^2) \right.
\nonumber \\ &&
 \left.+[b(x,\mu^2) + \bar b(x,\mu^2)] \theta(\mu^2 - m_b^2) 
+ g(x,\mu^2) \right]
 \,  = 1 \,.
\nonumber \\
\end{eqnarray}
Also the quark constituents carry all the charge, 
isospin, strangeness, charm and bottom quantum numbers of the nucleon
so they also satisfy the other standard sum rules for the conservation 
of these quantities, see \cite{rob}, \cite{esw}. 


%% file: my4.tex
\newcommand{\bea}{\begin{eqnarray}}
\newcommand{\eea}{\end{eqnarray}}

\mysection{Direct $x$-space method of solution and initial conditions}

\subsection{The method }

Our choice of the direct $x$-space method is motivated by the necessity 
to step densities across heavy flavor thresholds
using LO, NLO and NNLO boundary conditions.
The procedure of doing this with Mellin moments would involve taking 
moments of the densities and then inverting moments 
several times.
The direct $x$-space method is much more intuitive and
straightforward.
The main features of this method are linear interpolation over a grid
in $x$ and second-order interpolation over a grid in $t$. 
Let us describe it in more detail to point out where we 
differ from the method in \cite{botje}.

First we consider the $x$-variable in the evolution and write the 
right-hand-side of the evolution equation Eq.(\ref{eqn3.5}) for 
the non-singlet density as
\begin{equation}
\label{4.1}
I(x_{0})=\int
\frac{dx}{x} \frac{x_{0}}{x} P\left( \frac{x_{0}}{x}\right) 
q\left(x\right) \,, 
\end{equation}
where $x_0 \le x \le 1$ ,
\begin{equation}
\label{4.2}
q(x)=xf(x) \,,
\end{equation}
and
\begin{equation}
\label{4.3}
x_{0} < x_{1} <...< x_{n}<x_{n+1}\equiv 1 \,,
\end{equation}
with \( q(x_{n+1})=q(1)\equiv 0 \).
Between grid points $x_i$ and $x_{i+1}$, $x$ is chosen so that
\begin{equation}
\label{4.4}
q(x) =(1 - \bar x)q(x_{i}) + \bar x q(x_{i+1}) \,,
\end{equation}
with $\bar x = (x-x_{i})/(x_{i+1}-x_{i}) $.
Using this relation we convert the integral into a sum
\begin{equation}
\label{4.5}
I(x_{0})=\sum_{i=0}^{n+1} w(x_{i},x_{0})q(x_{i}) \,,
\end{equation}
where the weights are (in all orders LO, NLO and NNLO)
\begin{eqnarray}
\label{4.6}
w(x_{0},x_{0})&=&S_{1}(s_{1},s_{0}) 
\nonumber \\
w(x_{i},x_{0})&=&S_{1}(s_{i+1},s_{i})-S_{2}(s_{i},s_{i-1}) \,,
\end{eqnarray}
where $s_{i}=x_{0}/x_{i}$ and
\begin{eqnarray}
\label{4.7}
S_{1}(u,v)&=&\frac{v}{v-u}\int_u^v (z-u)P(z)\frac{dz}{z} \,,
\nonumber \\
S_{2}(u,v)&=&\frac{u}{v-u}\int_u^v (z-v)P(z)\frac{dz}{z} \,.
\end{eqnarray}

In the above formula $P(z)$ denotes the splitting function of 
the corresponding order in $\alpha_s$ and type (non-singlet, singlet, etc.)
We use the LO and NLO splitting functions in \cite{cfr} and  
the approximations to the NNLO splitting functions from \cite{3loop1} 
and \cite{3loop2}. For completeness the latter are given in Appendix A.
We have calculated the integrals in Eq.(\ref{4.7}) 
analytically and the results are in the computer program.  
This yields the formula in Eq.(\ref{4.5}) describing the grid 
for the $x$ variable. Note that the weights $w^{(0)}$, $w^{(1)}$ 
and $w^{(2)}$ are those for the exact LO, the exact NLO and the
approximate NNLO splitting functions respectively. 
Thus, for the singlet case, we have 
\begin{eqnarray}
\label{4.8}
\frac{d (x_{0}\Sigma (x_{0}))}{d t}
&=&\frac{\alpha_{s}}{2\pi}
\sum^{n+1}_{i=0}
\Big [ 
  [ 
w^{(0)}_{qq}(x_{i},x_{0})
+\frac{\alpha_{s}}{2\pi }w_{qq}^{(1)}(x_{i},x_{0})
+(\frac{\alpha_{s}}{2\pi })^2 w_{qq}^{(2)}(x_{i},x_{0})
 ]  
\nonumber \\ && \mbox{}
\times \, x_{i}\Sigma (x_{i})
\nonumber \\
&& +  [
w^{(0)}_{qg}(x_{i},x_{0})
+\frac{\alpha_{s}}{2\pi }w_{qg}^{(1)}(x_{i},x_{0})
+(\frac{\alpha_{s}}{2\pi })^2 w_{qg}^{(2)}(x_{i},x_{0})
 ] 
\nonumber  \\ && \mbox{}
\times \, x_{i} g(x_{i}) 
 \Big]
\,,
\end{eqnarray}
where $\Sigma$ is either $\Sigma(3)$, $\Sigma(4)$ or $\Sigma(5)$
depending on the scale.

Now consider the variation in the variable $t$. For each $x_i$ we 
pick a grid in $t$ labelled by distinct points $t_j$.  
Then, for example, the non-singlet equation becomes
\begin{eqnarray}
\label{4.9}
&& q^{'}(x_{i},t_{j})
=\frac{\alpha _{s}(t_{j})}{2\pi }
\sum_{k=1}^n \big[ w^{(0)}_{\pm }(x_{k},x_{i})
+\frac{\alpha_{s}(t_{j})}{2\pi }w^{(1)}_{\pm }(x_{k},x_{i})
\nonumber  \\ && \mbox{}
\qquad \qquad \qquad \qquad + \bigl(\frac{\alpha_{s}(t_j)}{2\pi }\bigr)^2 
w_{\pm}^{(2)}(x_{k},x_{i})\big]
q(x_{k},t_{j}) \,,
\end{eqnarray}
where $ q^{'}(x_{i},t_{j})$ denotes the derivative with respect to 
$t$ evaluated at $t = t_j$. 
In compact notation this equation can be rewritten as
\begin{equation}
\label{4.10}
q_{j}^{'}= wq_{j} + S \,,
\end{equation}
with $S$ being the sum of the terms on the right hand side 
of Eq.(\ref{4.9}) excluding the $j$-th term.

For $t$ between the grid points $t_{j-1}$ and $t_{j}$ we interpolate 
the parton density using quadratic interpolation as follows:
\begin{equation}
\label{4.11}
q(x_{i},t)=at^{2} + bt + c \,.
\end{equation}
Thus we relate the value of $q$ at the point $t_j$ to that of $q$
at the point $t_{j-1}$ by
\begin{equation}
\label{4.12}
q(x_{i},t_{j})=q(x_{i},t_{j-1})+\frac{1}{2}
[q^{'}(x_{i},t_{j})+q^{'}(x_{i},t_{j-1})]\Delta t_{j} \,,
\end{equation}
where \( \Delta t_{j}=t_{j}-t_{j-1} \). This equation can also be written 
more compactly as
\begin{equation}
\label{4.13}
q_j=q_{j-1}+\frac{1}{2}(q^{'}_{j-1}+q^{'}_{j})\Delta t_j \,.
\end{equation}
The resulting system of two linear equations in Eq.(\ref{4.10}) 
and Eq. (\ref{4.13}) for $q_j$ and $q^{'}_j$ has the solution 
\begin{equation}
\label{4.14}
q_j=\frac{2q_{j-1}+(q^{'}_{j-1}+S)\Delta t_j}{2-w\Delta t_j} \,.
\end{equation}
Then we find $q^{'}_j$ from Eq.(\ref{4.10}). 
Applying the same procedure to the gluon and singlet combinations 
involves four equations because we have to compute both the 
densities and their derivatives.

The evolution proceeds from the initial $\mu_0^2 = \mu^2_{\rm LO}$ (or 
$\mu_0^2 = \mu^2_{\rm NLO}$) to the first heavy flavor threshold
at the scale $\mu^2 = m_c^2$. Next the charm density is introduced in 
NNLO ($\alpha^{2}_{s}$-order terms) and all the four-flavor
densities are evolved from the new boundary conditions in 
Section 4.2.
This evolution continues up to the  
transition point $\mu^2 = m_b^2$, where the same procedure 
is applied to generate the bottom quark density. From that 
matching point all five-flavor densities are evolved 
up to all higher $\mu^2$ scales starting from the 
boundary conditions in Appendix B.

Since the weights for the calculation are computed analytically from the
LO, NLO \cite{cfr} and NNLO (\cite{3loop1},\cite{3loop2}) $\overline{\rm MS}$
splitting functions we remove possible
instabilities in the numerical integrations. 
Hence the program is very efficient and fast.
The results from the evolution code have been 
thoroughly checked against the tables in the HERA report \cite{brnv}
and they agree to all five decimal places.

\subsection{The initial conditions}

The GRV98 \cite{grv98} three-flavor LO and NLO parton density sets 
contain input formulae at low scales $\mu < m_c$ which are 
ideal as initial values for our parametrizations. Therefore 
we start our LO evolution using the following input 
at $\mu_0^2=\mu_{\rm LO}^2=0.26$ ${\rm GeV}^2$ 
\begin{eqnarray}
\label{4.15}
x f_{u-\bar u}(3,x,\mu_0^2)& = &
xu_v(x,\mu_{\rm LO}^2) \nonumber\\
& = & 1.239\,\,x^{0.48}\,(1-x)^{2.72}\, 
(1-1.8\sqrt{x} + 9.5x) \nonumber\\
x f_{d-\bar d}(3,x,\mu_0^2) & = &    
xd_v(x,\mu_{\rm LO}^2) \nonumber\\
& = & 0.614\,\, (1-x)^{0.9}\,\, 
xu_v(x,\mu_{\rm LO}^2)\nonumber\\
x (f_{\bar d}(3,x,\mu_0^2)-f_{\bar u}(3,x,\mu_0^2)) & = &
x\Delta(x,\mu_{\rm LO}^2) \nonumber \\
& = & 0.23 \,\, x^{0.48}\,(1-x)^{11.3}\,
(1-12.0\sqrt{x} + 50.9x)\nonumber\\
x (f_{\bar d}(3,x,\mu_0^2)+f_{\bar u}(3,x,\mu_0^2)) & = &
x(\bar{u}+\bar{d})(x,\mu_{\rm LO}^2) \nonumber \\
& = & 1.52\,\, x^{0.15}\, (1-x)^{9.1}\,
(1-3.6\sqrt{x} + 7.8x)\nonumber\\
x f_g(3,x,\mu_0^2) & = & 
xg(x,\mu_{\rm LO}^2) \nonumber \\
 & = & 17.47\,\, x^{1.6}\, (1-x)^{3.8}\nonumber\\
x f_{s}(3,x,\mu_0^2) =   
x f_{\bar s}(3,x,\mu_0^2)  & = &  xs(x,\mu_{\rm LO}^2) \nonumber \\  
& = & x\bar{s}(x,\mu_{\rm LO}^2) = 0 \,.
\end{eqnarray}
Here $\Delta\equiv\bar{d}-\bar{u}$ is used to construct 
the non-singlet combination.

We start the corresponding NLO evolution using the following 
GRV98 input at $\mu_0^2=\mu_{\rm NLO}^2=0.40$ ${\rm GeV}^2$ 
\begin{eqnarray}
\label{4.16}
x f_{u-\bar u}(3,x,\mu_0^2) & = & 
xu_v(x,\mu_{\rm NLO}^2) \nonumber \\
& = & 0.632\,\,x^{0.43}\,(1-x)^{3.09}\, 
(1+18.2x)\nonumber\\
x f_{d-\bar d}(3,x,\mu_0^2) & = &    
xd_v(x,\mu_{\rm NLO}^2) \nonumber \\
& = & 0.624\,\, (1-x)^{1.0}\,\, 
xu_v(x,\mu_{\rm NLO}^2)\nonumber\\
x (f_{\bar d}(3,x,\mu_0^2)-f_{\bar u}(3,x,\mu_0^2)) & = &
x\Delta(x,\mu_{\rm NLO}^2) \nonumber \\
& = & 0.20 \,\, x^{0.43}\,(1-x)^{12.4}\,
(1-13.3\sqrt{x} + 60.0x)\nonumber\\
x (f_{\bar d}(3,x,\mu_0^2)+f_{\bar u}(3,x,\mu_0^2)) & = &
x(\bar{u}+\bar{d})(x,\mu_{\rm NLO}^2) \nonumber \\
& = & 1.24\,\, x^{0.20}\, (1-x)^{8.5}\,
(1-2.3\sqrt{x} + 5.7x)\nonumber\\
x f_g(3,x,\mu_0^2) & = & xg(x,\mu_{\rm NLO}^2) \nonumber \\ 
& = & 20.80\,\, x^{1.6}\, (1-x)^{4.1}\nonumber\\
x f_{s}(3,x,\mu_0^2) =   
x f_{\bar s}(3,x,\mu_0^2) & = &
xs(x,\mu_{\rm NLO}^2)\nonumber \\
 & = & x\bar{s}(x,\mu_{\rm NLO}^2) = 0.
\end{eqnarray}

We start the corresponding NNLO evolution using the 
same NLO input and starting scale as above.

\subsection{The calculation of the running coupling}

The heavy quark masses $m_c = 1.4$ ${\rm GeV}^2$, $m_b = 4.5$
$ {\rm GeV}^2$ are used throughout the calculation. 
We also use the exact solution (as opposed to a perturbative solution 
in inverse powers of $\ln (\mu^2/\Lambda^2)$)
of the differential equation
\begin{equation}
\label{4.17}
\frac{d\, \alpha_s(\mu^2)}{d\, \ln(\mu^2)}= 
-\frac{\beta_0}{4\pi}\, \alpha_s^2(\mu^2)
\, -\, \frac{\beta_1}{16\pi^2}\, \alpha_s^3(\mu^2) \,,
\end{equation}
for the running coupling $\alpha_s(\mu^2)$.
Here $\beta_0=11 - 2n_f/3$ and $\beta_1=102 - 38n_f/3$.
Another way of writing this equation is 
\begin{equation}
\label{4.18}
\ln\frac{\mu^2}{(\tilde{\Lambda }^{(n_f)}_{\rm EXACT})^2} 
= \frac{4\pi} {\beta_0\alpha_s(\mu^2)}\,-\, \frac{\beta_1}{\beta_0^2} \, \ln
\left[ \frac{4\pi}{\beta_0\alpha_s(\mu^2)}\, +\, \frac{\beta_1}{\beta_0^2}
\right] \,.
\end{equation}
The values for $\tilde{\Lambda}^{(n_f)}_{\rm EXACT}$
are carefully chosen to obtain accurate matching of $\alpha_s$ 
at the scales $m_c^2$ and $m_b^2$. We used the values 
$\tilde {\Lambda}^{(3,4,5,6)}_{\rm EXACT}=299.4,$\,
$246,\, 167.7,$ $\, 67.8$ 
MeV$/c^2$
respectively in the exact formula 
(which yields 
$\alpha_s^{\rm EXACT}(m_Z^2)$ $= 0.114$,
$\alpha_s^{\rm EXACT}(m_b^2)$ $= 0.205$,
$\alpha_s^{\rm EXACT}(m_c^2)$ $= 0.319$,
$\alpha_s^{\rm EXACT}(\mu_{\rm NLO}^2)$ $= 0.578$
) and
$\Lambda_{\rm LO}^{(3,4,5,6)} = 204,$\,
$175,\, 132,$\,$ 66.5$ MeV$/c^2$ 
respectively 
(which yields 
$\alpha_s^{\rm LO}(m_Z^2)$ $= 0.125$,
$\alpha_s^{\rm LO}(m_b^2)$ $= 0.232$,
$\alpha_s^{\rm LO}(m_c^2)$ $= 0.362$,
$\alpha_s^{\rm LO}(\mu_{\rm LO}^2)$ $= 0.763$
) 
for the LO formula (where $\beta_1=0$). 
There is a NNLO discontinuity of aproximately two parts in one
thousand in the running coupling across heavy flavor thresholds 
\cite{2loop}, \cite{2loop2}. We have ignored this effect
to focus on the numerically more significant 
matching of the flavor densities. 

\subsection{The evolution process}

Three flavor evolution proceeds from the initial $\mu_0^2$ to the 
scale $\mu^2=m_c^2=1.96$ $({\rm GeV}^2)^2$. 
At this point the charm density is then defined by 
\begin{eqnarray}
\label{4.19}
f_{c + {\bar c}}(n_f+1,m_c^2)&=&
a_s^2 (n_f,m_c^2)\Big [\tilde A_{Qq}^{\rm PS}
(1)\otimes f_q^{\rm S}(n_f, m_c^2)
\nonumber\\[2ex]  
&& + \tilde A_{Qg}^{\rm S}(1) 
\otimes f_g^{\rm S}(n_f, m_c^2)\Big ] \,,
\end{eqnarray}
with $n_f=3$ and $a_s = \alpha_s/4\pi$.
We have suppressed the $x$ dependence to make the notation more compact.
The $\otimes$ symbol denotes the convolution integral
$f\otimes g=\int f(x/y)g(y)dy/y$, where $x \le y \le 1$.
The OME's $\tilde A^{\rm PS}_{Qq}(\mu^2/m_c^2)$,
$\tilde A^{\rm S}_{Qg}(\mu^2/m_c^2)$
are given in \cite{bmsn1} and are also listed in Appendix B. 
The reason for choosing the matching scale $\mu$ 
at the mass of the charm quark $m_c$ is that all the $\ln(\mu^2/m_c^2)$ 
terms in the OME's vanish at this point leaving only the
nonlogarithmic pieces in the order $\alpha_s^2$ OME's to contribute to
the right-hand-side of Eq.(4.19). Hence the LO and NLO charm densities
vanish at the scale $\mu=m_c$. The NNLO charm density starts off with 
a finite $x$-dependent shape in order $\alpha_s^2$. 
Note that we then order the terms on the right-hand-side of
Eq. (\ref{4.19}) so that it contains a product of NLO OME's and 
LO parton densities. The result is then of order $\alpha_s^2$ 
and should be multiplied by order $\alpha_s^0$ coefficient 
functions when forming the deep inelastic structure functions. 

The four-flavor gluon density is also generated at the
matching point in the same way. At $\mu = m_c$ we define
\begin{eqnarray}
\label{4.20}
f_g^{\rm S}(n_f+1, m_c^2) &=& f_g^{\rm S}(n_f, m_c^2)
\nonumber\\
&&  + a_s^2(n_f,m_c^2) \Big [ A_{gq,Q}^{\rm S}(1)
\otimes f_q^{\rm S}(n_f, m_c^2) \,, 
\nonumber\\
&&+ A_{gg,Q}^{\rm S}(1)
\otimes f_g^{\rm S}(n_f, m_c^2) \Big ]\,.
\end{eqnarray}
The OME's $A^{\rm S}_{gq,Q}(\mu^2/m_c^2)$, $A^{\rm S}_{gg,Q}(\mu^2/m_c^2)$
are given in \cite{bmsn1} and are also listed in the Appendix B. 
The four-flavor light quark (u,d,s) densities are generated using
\begin{eqnarray}
\label{4.21}
f_{k+\bar k}^{\rm}(n_f+1,m_c^2) &=& f_{k+\bar k}(n_f,m_c^2) 
\nonumber\\
&& + a_s^2(n_f,m_c^2) A_{qq,Q}^{\rm NS}(1)
\otimes f_{k+\bar k}(n_f, m_c^2)\,.
\end{eqnarray}
The OME $A^{\rm NS}_{qq,Q}(\mu^2/m_c^2)$ is given in \cite{bmsn1} 
(as well as in Appendix B)
and the {\it total} four-flavor singlet quark density 
folows from the sum of Eqs. (\ref{4.19}) and (\ref{4.21}).
In Eqs. (\ref{4.20}) and (\ref{4.21}) we set $n_f = 3$. The remarks 
after Eq. (\ref{4.19}) are relevant here too.

Next the resulting four-flavor densities are evolved using the 
four-flavor weights
in either LO, NLO and NNLO up to the scale
$\mu^2 = m_b^2 = 20.25$ $({\rm GeV}^2)^2$. 
The bottom quark density is then generated at this point using 
\begin{eqnarray}
\label{4.22}
f_{b + {\bar b}}(n_f+1,m_b^2)&=&
a_s^2 (n_f,m_b^2)\Big [\tilde A_{Qq}^{\rm PS}
(1)\otimes f_q^{\rm S}(n_f, m_b^2)
\nonumber\\
&&+ \tilde A_{Qg}^{(\rm S)}(1) 
\otimes f_g^{\rm S}(n_f, m_b^2)\Big ] \,,
\end{eqnarray}
and the gluon and light quark densities (which now include charm) 
are generated using
Eqs.(\ref{4.19})-(\ref{4.21}) 
with $n_f =4$ and replacing $m_c^2 $ by $m_b^2$. 
Therefore only the nonlogarithmic terms in the order $a_s^2$ OME's 
contribute to the matching conditions on the bottom quark density.
Then all the densities are evolved up to higher $\mu^2$ as a five-flavor set
with either LO, NLO and NNLO splitting functions. This is valid until
$\mu = m_t \approx 175$ ${\rm GeV}^2$ above which one should switch
to a six-flavor set. We do not implement this step because the top
quark density would be extremely small.

The procedure outlined above generates a full set of parton densities
(gluon, singlet, non-singlet light and heavy quark densities,) 
for any $x$ and $\mu^2$ from the three-flavor LO, 
NLO and NNLO inputs in Eqs.(\ref{4.15}) and (\ref{4.16}).
Note that we have used $\mu^2$ for the factorization and
renormalization scales in the above discussion. In the computer program we
use the notation that $Q^2$ denotes these scales, since this is 
done in all the previous computer codes for the parton densities.



%% file: my5.tex
\mysection{Input parameter description and usage }

To prepare the program for use unpack the distribution package
{\bf adens-24.tar.gz} by typing tar -xzf adens-24.tar.gz. 
The resulting directory will contain the following files

\begin{verbatim}
head.h
main.h
main.c
l-a-w.c
nl-a-w.c
alpha.c
init.c
polylo.c
intpol.c
evolver.c
thresh.c
a-coefs.c
loader.c
quadrat.c
daind.c
integrands.c
grids.c
weights.c
nnl-a-w.c
wgplg.c
evolution_parameters.input
makefile
my_howto.tex
sample.out

\end{verbatim}

To build the executable on a machine with a gcc compiler
type {\it make }. The executable named
{\bf adens.x} will be produced.
To run the code just run the file {\bf adens.x}. Some debugging information
may appear on the standard output.

Here is the parameter file ({\bf evolution\_parameters.input}) 
explanation with default values shown:

\begin{tabular}{|r|l|p{6cm}|}
\hline
0.204e0    &     LambdaLO-3          &            LO $\Lambda$ for $N_f=$3 \\
\hline
0.175e0    &       LambdaLO-4       &               LO $\Lambda$ for $N_f=$4 \\
\hline
0.132e0    &       LambdaLO-5        &              LO $\Lambda$ for $N_f=$5 \\
\hline
0.306e0    &       LambdaNLO3       &          NLO $\Lambda$ for $N_f=$3     \\
\hline
0.257e0    &       LambdaNLO4       &            NLO $\Lambda$ for $N_f=$4 \\
\hline
0.1734e0    &      LambdaNLO5       &            NLO $\Lambda$ for $N_f=$5 \\
\hline
0.2994e0   &       LambdaENLO3       &           Exact $\Lambda$ for $N_f=$3 \\
\hline
0.246e0     &      LambdaENLO4       &           Exact $\Lambda$ for $N_f=$3 \\
\hline
0.1677e0    &      LambdaENLO5       &           Exact $\Lambda$ for $N_f=$3 \\
\hline
0.40e0      &      Qinitial2         &       Initial $Q^2$ to start evolution \\
\hline
1.96e0     &       QcharmMass       &           Mass of first heavy quark c \\
\hline
20.25e0    &       QbottomMass       &         Mass of second heavy quark b \\
\hline
1.96e0     &       QcharmThreshold      &            Charm threshold \\
\hline
1.96e0    &        AlphaCharmThreshold   &    C threshold used for $\alpha_s$ \\
\hline
20.25e0    &       QbottomThreshold        &         Bottom threshold \\
\hline
20.25e0    &       AlphaQbottomThreshold    &   B threshold used for $\alpha_s$ \\
\hline
1000.0e0  &        Qfinal2              &            Final $Q^2$ \\
\hline
130      &         tGridSize      &                 $Q^2$ grid size \\
\hline
200       &        xGridSize     &                  $x$ grid size \\
\hline
130       &        xGridSplit     &        $x$ split between log and linear \\
\hline
1.0e-5    &        xInitial     &                    $x$ initial \\
\hline
0.2e0     &        xSplit      &        $x$ at the split btw log and linear \\
\hline
1.00e0    &        xFinal      &                     $x$ final (always 1)  \\
\hline
0        &         DebugLevel     &         Error message detail (0-5)    \\
\hline
1        &         GraphVsX      &           Plotting data files are versus
                                               $x$ (1) or $Q^2$ (0) \\
\hline
1        &         Order      &                   LO/NLO/NNLO for 0,1,2 \\
\hline
0        &         DoFortran      &               Produce (1) or no (0) 
                                               data files for CSN/BMSN Fortran 
                                               programs (1-yes, 0-no) \\
\hline
1&AlphaDoSeparateThreshold&Use separate thresholds for $\alpha_s$ (1-yes, 0-no) \\
\hline
\end{tabular}  

\begin{tabular}{|r|l|p{6cm}|}
\hline
1        &  AlphaUseExact  &  Use exact GRV98-style $\alpha_s$ (1-yes, 0-no) \\
\hline
0        &         ThreeFlavorMode   &      Calculate GRV98-style densities 
                                         with no heavy flavors (1-yes, 0-no) \\
\hline
0        &         GraphAll      &     Plot all data points (1-yes, 0-no) \\
\hline
0         &        NNLOmultiOrderCHARM    &          Use our proper order NNLO 
                                               heavy flavors (1-yes, 0-no) \\
\hline
1         &        DoBottomThreshold     &    Generate bottom (1-yes, 0-no) \\
\hline
0        &         LoadWeightsMadeBefore   &         Use ready weights if 
                                               available (1-yes, 0-no) \\
\hline
1        &         DoNotDumpWeights     &       Dump weight for future use as 
                                              the option above (1-yes, 0-no) \\
\hline
0        & NLO4NNLO  & Use NLO weights for NNLO calculation (1-yes, 0-no)  \\
\hline
\end{tabular}  

\medskip 
The first set of Lambdas are used for LO
calculations. The second set are used for NLO and NNLO calculations if
the exact $\alpha_s$ is not requested (AlphaUseExact=0). 
The next set (LambdaENLO3, LambdaENLO4, LambdaENLO5 ) 
are used for the exact solution of the differential equation for 
$\alpha_s$ as proposed in the GRV98 paper \cite{grv98}. 
The code that calculates the exact $\alpha_s$ might 
use its own set of flavor thresholds (which means that the number
of flavors used for $\alpha_s$ can be reset independently from the regular 
heavy flavor threshold as done in \cite{grv98}).

Next we give the $Q^2$ limits and the heavy masses: the 
initial and final $Q^2$, the charm and bottom masses 
(used in threshold calculations), the
heavy flavor thresholds and the separate $\alpha_s$ thresholds.  
Next follow the grid sizes in $x$ and $Q^2$ together 
with $x$ initial and final (always $1$) and the switch point between
logarithmic and linear grids in the $x$ dimension.  
The $x$ grid always starts as logarithmic 
and then becomes linear at higher $x$, usually at
a value of the order of 0.1 (xGridSplit parameter).

The last group of parameters contains various control 
values that set the modes of the computation: \\
{\it DebugLevel }, controls the amount of generated error, warning and
information messages, \\
{\it GraphVsX }, controls the printing of the output data for plotting 
(first column is either $x$ or $Q^2$, then subsequent columns
will contain density values for various  $Q^2$ or $x$), \\  
{\it Order}, sets calculation order (use 0,1,2 for LO,NLO,NNLO),\\
{\it DoFortran}, sets  whether to dump interpolated densities 
on a special grid for future use in Fortran code for the
calculation of structure functions ; CSN and BMSN refer to VFNS schemes
which are explained in \cite{csn}, \\
{\it AlphaDoSeparateThreshold}, sets  whether we use a separate threshold for
$\alpha_s$ (used, for instance for GRV98 set where $n_f$ for densities is 
always 3 and $n_f$ for  $\alpha_s$ goes from 3 to 5,\\
{\it AlphaUseExact}, sets whether to use exact (differential equation
solution) $\alpha_s$ for NLO and NNLO calculation,\\
{\it ThreeFlavorMode}, sets whether to run GRV98 mode (no heavy flavors,
$n_f=3$ for all $Q^2$),\\
{\it GraphAll}, controls the amount of graphing and printing output 
(either all data points or the
special grid defined in the file {\bf main.h}, 
that contains some favorite values (for more see Section 7)),\\
{\it NNLOmultiOrderCHARM}, activates NNLO threshold calculation using proper
order combinations (this mode requires one to first run the LO and NLO 
calculations),\\
{\it DoBottomThreshold}, enables the bottom density,\\
{\it LoadWeightsMadeBefore}, turns on and off the loading of weights 
computed in the prior runs,\\
{\it DoNotDumpWeights}, sets whether to save computed weights to disk for
future use,\\
{\it NLO4NNLO}, sets whether NLO weights are used for the NNLO
calculation (thus having only the boundary condition in NNLO).

Some common parameter settings and typical grid sizes for 
popular evolutions are shown in Section 7.


%% file: my6.tex
\mysection{Description of the program}

\subsection{Program module summary}

\begin{tabular}{|r|l|}
\hline
main.c        &           The main program, input and output      \\  
l-a-w.c       &           Calculation  of LO weights    \\
nl-a-w.c      &           Calculation of NLO weights    \\
alpha.c       &           Calculation of $\alpha_s$    \\
init.c        &           Definition of initial functions     \\
polylo.c      &           Calculation of  polylogarithms    \\
intpol.c      &           Interpolation routine     \\
evolver.c     &           Evolution process subroutine     \\
thresh.c      &           Threshold handling subroutine     \\
a-coefs.c    &           OMEs for thresholds     \\
loader.c     &           Datafile reading subroutine     \\
quadrat.c    &           Gaussian integration subroutine     \\
daind.c      &           Another integration subroutine     \\
integrands.c &           Heavy flavor integrand calculation routine   \\
grids.c      &           Grid generation routine and            \\
{}               &           memory management routines   \\
weights.c    &           Weight table handling routine     \\
nnl-a-w.c    &           Calculation of NNLO weights    \\
wgplg.c      &           Calculation of  high order polylogarithms    \\
\hline
\end{tabular}

\subsection{main.c} 

{\it subroutines:} \\ 
none. \\

The main program module contains input handling from 
the parameter file, parameter
verification, calls to grid generating routines 
({\bf MakeXGrid, MakeTGrid}),
resets for all density arrays (array q) and their derivatives (array qp).
It also includes calls to the generation of 
weights ({\bf analowgts, ananlowgts }),
the calls to evolution and threshold routines ({\bf  evolver,
threshold}) that do the actual work. 
Also it contains some pre-output density processing
and the results provided in various formats for both viewing and plotting.

\subsection{l-a-w.c}

{\it subroutines:}  \\
int analowgts(int nf,int loadWgts), \\
int computeLOwgts(int nf), \\
double sqq(double x,double y), \\
double sgg(double x,double y).\\

Analytically computes or reads from the file the LO weights 
for the evolution equations.

\subsection{nl-a-w.c}

{\it subroutines:} \\ 
int ananlowgts(int nf,int loadWgts),\\
int computeNLOwgts(int nf),\\
    double s1ff(double x,double y, int nf),\\
    double s2ff(double x,double y, int nf),\\
    double s1fg(double x,double y, int nf),\\    
    double s2fg(double x,double y, int nf),\\ 
    double s1gf(double x,double y, int nf),\\    
    double s2gf(double x,double y, int nf),\\ 
    double s1gg(double x,double y, int nf),\\    
    double s2gg(double x,double y, int nf),\\ 
    double s1ff\_plus(double x, int nf),\\        
    double s1gg\_plus(double x, int nf),\\
    double s1p(double x,double y, int nf),\\     
    double s2p(double x,double y, int nf),\\ 
    double s1m(double x,double y, int nf),\\     
    double s2m(double x,double y, int nf),\\ 
    double s1p\_plus(double x, int nf),\\         
    double s1m\_plus(double x, int nf),\\
    double s1gf\_lim(double sp,double nf),\\      
    double s1fg\_lim(double sp,double nf),\\ 
    double s2ff\_lim(double sp,double nf),\\      
    double s2fg\_lim(double sp,double nf),\\ 
    double s2gf\_lim(double sp,double nf),\\      
    double s2gg\_lim(double sp,double nf),\\
    double s2p\_lim(double sp,double nf),\\       
    double s2m\_lim(double sp,double nf).\\

Analytically computes or reads from the file the NLO weights for the
evolution equation. These routines are grouped into 3 kinds: the
s{1,2}xx routines calculate the regular weights, the s{1,2}xx\_lim routines
calculate the regular weights called at 
1 and s{1,2}xx\_plus do the weights that
contain the plus-distributions.

\subsection{alpha.c}

{\it subroutines:} \\
double alpha(double tt, int nf),
double alphae (double tt,int nf). \\

Calculates LO, NLO and exact running coupling 
$\alpha_s$ using corresponding
parameters from the input file.

\subsection{init.c}

{\it subroutines:}\\
double initq\_uv(double xx),\\
double initq\_dv(double xx),\\
double init\_gl(double xx),\\
double initq\_ss(double xx),\\
double initq\_del(double xx),\\
double initq\_udbar(double xx).\\
\\

Sets initial values for all parton densities 
using the GRV98 input for LO and NLO densities from \cite{grv98}.

\subsection{polylo.c}

{\it subroutines:} \\
double Li2(double x),\\
double Li3(double x),\\
double S12(double x).\\ 
\\

Calculates these three polylogarithms
using a fast routine with Bernouilli numbers.
.

\subsection{intpol.c}

{\it subroutines:} \\
double int\_q(int j,double xx,int it),\\
double interpolate(double xx,double *xt, double *yt,int points).
\\

Interpolation routines used to calculate densities between 
grid points and for integration at the threshold.

\subsection{evolver.c}

{\it subroutines:} \\
evolver(int it1,int it2,int ic,int ib).\\

The main routine that performs the evolution between thresholds 
for all densities. It
updates the main density array q and the density derivatives array qp.

\subsection{thresh.c}

{\it subroutines:} \\
int threshold(int what,int itt),\\
int fdens4(double  xx,int ittc,double *u,double *d,double *s),\\
double light\_charm(double  xx,int ittc),\\
double  fcharm(double xx,int ittc),\\
double  fbottom(double xx,int ittc),\\
double  fsigma(double xx,int ittc),\\
double  fgluon(double xx,int ittc),\\
double  fcharm(double xx,int ittc),\\
double  fbottom(double xx,int ittc).\\
\\

Threshold handling routines to implement LO, NLO and NNLO matching 
conditions for light and
heavy densities at the charm and bottom thresholds. 
The density routines are calls
to convolution integrals that generate new densities for $n_f+1$ flavors.

\subsection{a-coefs.c }

{\it subroutines:}\\
double  a1qg(double z,double fs2,double hm2),\\
double  a2qq(double z,double fs2,double hm2),\\
double  a2qg(double z,double fs2,double hm2),\\
double  a2qqns(double z,double fs2,double hm2),\\
double  softq(double z,double fs2,double hm2),\\
double  corq(double z,double fs2,double hm2),\\
double  a2gg(double z,double fs2,double hm2),\\
double  softg(double z,double fs2,double hm2),\\
double  corg1(double fs2,double hm2),\\
double  corg2(double z,double fs2,double hm2),\\
double  a2gq(double z,double fs2,double hm2).\\
\\

The OME routines used for NNLO threshold matching. These contain
the formulae in Appendix B.

\subsection{loader.c}

{\it subroutines:} \\
int loadOrd(int what).\\

Functions to handle threshold datafile loading, saving and verification. This
file allows one the ability to use previously computed density values at the 
threshold in a new computation.

\subsection{quadrat.c }
  
{\it subroutines:} \\
double qadrat(double *x, double a, double b, double (*fx)(double), double e[]),
\\
double lint(double *x, double (*fx)(double), double e[], double x0,
double xn, double f0, double f2, double f3, double f5,
double f6, double f7, double f9, double f14,
double hmin, double hmax, double re, double ae).
\\

Backup integration routine used as a check for the actual one 
used in the threshold integration.

\subsection{daind.c}

{\it subroutines:} \\
double daind(double  *x,double  a,double b, 
double (*fun)(double),double eps,int key,int max).
\\

Main Gaussian integration routine, see 
\cite{pie}.   

\subsection{integrands.c}

{\it subroutines:}\\
inline double  fcharm\_integrand(double  x1),\\
inline double  fgluon\_integrand(double x1),\\
inline double  fsigma\_integrand(double x1),\\
inline double us\_integrand(double  x1),\\
inline double ds\_integrand(double  x1),\\
inline double  ss\_integrand(double  x1),\\
inline double  fbottom\_integrand(double  x1),\\
inline double light\_charm\_integrand(double  x1).\\
\\

Functions containing integrands for the threshold integration. They use the
density values and the coefficient functions from a-coefs.c to produce the
integrands that are then fed into the Gaussian integration program.

\subsection{grids.c}

{\it subroutines:} \\
int MakeXGrid(void),\\
int MakeTGrid(void),\\
int merge(double  *a,double  *b,int na, int nb,char w),\\
int check\_grid(double *a,int n,char w),\\
int MakeFortranGrid(int test\_mode),\\
double **allocate\_real\_matrix(int ur, int uc),\\
void free\_real\_matrix(double **m,int ur).\\
\\

Subroutines for making (and also merging and verifying)
the initial grids in $x$ and $Q^2$ and the final grids for
Fortran-code compatible output. The grid merging is used to combine the evenly
spaced grid generated automatically from the initial and final values with the
premade grid containing several $x$ and $Q^2$ values for plotting and
outputting the data. Two routines are added for deallocating memory.

\subsection{weights.c}

{\it subroutines:}\\
int readWeights(int nf,int order),\\
int dumpWeights(int nf,int order).\\

Routines dealing with loading and saving computed NLO and NNLO 
weight tables to do a
fast calculation on the same grids. LO weights are not saved as it is very
fast to compute them every time.

\subsection{nnl-a-w.c}

{\it subroutines:}\\
int anannlowgts(int nf,int loadWgts),\\
int computeNNLOwgts(int nf),\\ 
    double nn\_s1ff(double x,double y, int nf),\\
    double nn\_s2ff(double x,double y, int nf),\\
    double nn\_s1fg(double x,double y, int nf),\\    
    double nn\_s2fg(double x,double y, int nf),\\ 
    double nn\_s1gf(double x,double y, int nf),\\    
    double nn\_s2gf(double x,double y, int nf),\\ 
    double nn\_s1gg(double x,double y, int nf),\\    
    double nn\_s2gg(double x,double y, int nf),\\ 
    double nn\_s1ff\_plus(double x, int nf),\\        
    double nn\_s1gg\_plus(double x, int nf),\\
    double nn\_s1p(double x,double y, int nf),\\     
    double nn\_s2p(double x,double y, int nf),\\
    double nn\_s1m(double x,double y, int nf),\\     
    double nn\_s2m(double x,double y, int nf),\\
    double nn\_s1p\_plus(double x, int nf),\\         
    double nn\_s1m\_plus(double x, int nf),\\
    double nn\_s1gf\_lim(double sp,double nf),\\      
    double nn\_s1fg\_lim(double sp,double nf),\\ 
    double nn\_s2ff\_lim(double sp,double nf),\\      
    double nn\_s2fg\_lim(double sp,double nf),\\ 
    double nn\_s2gf\_lim(double sp,double nf),\\      
    double nn\_s2gg\_lim(double sp,double nf),\\
    double nn\_s2p\_lim(double sp,double nf),\\       
    double nn\_s2m\_lim(double sp,double nf).\\

Analytically computes or reads from files 
the approximate NNLO weights for the evolution equations. 
Here the routines are grouped into three kinds: the
nn\_s{1,2}xx routines calculate the regular weights, 
the nn\_s{1,2}xx\_lim routines calculate the regular weights 
called at 1 and nn\_s{1,2}xx\_plus do the weights that
contain the plus-distributions.
   
\subsection{wgplg.c}

{\it subroutines:}\\
double  wgplg(int n,int p,double  x).

The routines which calculate polylogarithms using the method
from CERNLIB \cite{poly}. They are only used for the higher order 
polylogarithms because the routines for Li2, Li3 and S12 
in polylo.c are faster.
\\


%% file: my7.tex
\mysection{Results}

The code can be used in several modes of operation.

For all of them there is some optimum grid size in $x$ and $Q^2$.
Internally, the grid with the sizes entered in the parameter file is merged
with another grid (that is used for plotting the output data at the end), thus
increasing the resulting grid size. This internal grid size contains all
``popular'' values, like $x= 0.1$, $0.01$, $0.001$ etc., and 
is 38 in $Q^2$ and 64 in $x$. 
The corresponding values are located in file {\bf main.h} 
(arrays xpr[] and q2pr[]). 
This grid is then merged with the automatically generated
equidistant grid and the equal values are weeded out. Shown in the table are
the resulting grid sizes as shown in the output file. The table uses the
calculation for all flavors  as opposed to the GRV98-like (only three-flavor)
densities. In general, the evolution
time grows quadratically in ${n_x}$ and linearly in $n_{Q^2}$. The numbers
we give below are for an alpha PC with a 21164 processor unit 
running at 500 MHz, 1 Gbyte of memory and rated at an Specfp = 20.4.
\vskip 0.8cm

\begin{tabular}{|r|l|l|l|l|}
\hline
order & $n_x$   & $n_{Q^2}$    & accuracy,digits     & time,sec \\
\hline
\hline
LO & 162   & 96    & 5     & 15 \\
\hline
NLO & 162   & 96    & 3     & 113 \\
\hline
NNLO & 162   & 96    & 3     & 385 \\
\hline
LO & 262   & 136    & 6     & 31 \\
\hline
NLO & 262   & 136    & 5     &  275 \\
\hline
NNLO & 262   & 136    & 5     & 1021 \\
\hline
LO & 362   & 136    & 6     & 44 \\
\hline
NLO & 362   & 136    & 6     &  529 \\
\hline
NNLO & 362   & 136    & 5     & 1537 \\
\hline
\end{tabular}
\vskip 0.8cm

1. Set parameters to the following values to produce LO and NLO GRV98-style
fixed three-flavor densities for the whole range of $Q^2$ (only parameters
essential for this calculation are provided, the rest can be set to whatever
one wishes since they control the form of the output and similar features, 
not the physically meaningful ones):
\vskip 0.8cm

\begin{tabular}{|r|l|}
\hline
LO & \space \\
\hline
0.26 & Qinitial2 \\
\hline
0 & Order \\
\hline
1 & ThreeFlavorMode \\
\hline \hline
NLO & \space \\
\hline
0.40 &         Qinitial2  \\
\hline
1      &         Order  \\
\hline
1       &        AlphaUseExact   \\
\hline
1       &        ThreeFlavorMode  \\
\hline
\end{tabular}  
\vskip 0.8cm

2. To generate regular VFNS densities with all heavy flavors (both charm and
bottom) one sets:
\vskip 0.8cm
\begin{tabular}{|r|l|}
\hline
LO & \space \\
\hline
0.26 & Qinitial2 \\
\hline
0 & Order \\
\hline
0 & ThreeFlavorMode \\
\hline
1      &           AlphaDoSeparateThreshold  \\
\hline
0        &         NNLOmultiOrderCHARM \\
\hline
1        &         DoBottomThreshold \\
\hline \hline
NLO & \space \\
\hline
0.40 &         Qinitial2  \\
\hline
1      &         Order  \\
\hline
0       &        ThreeFlavorMode  \\
\hline
1        &         AlphaDoSeparateThreshold    \\
\hline
1        &         AlphaUseExact   \\
\hline
0        &         NNLOmultiOrderCHARM   \\
\hline
1         &        DoBottomThreshold   \\
\hline \hline
\end{tabular}
\vskip 0.8cm
3. To generate VFNS densities involving proper order mixing at heavy thresholds
with all heavy flavors (both charm and bottom) but without using NNLO weights 
(as done in our previous papers \cite{csn}, \cite{cbot},
\cite{csh}, \cite{cs}) one sets:
\vskip 0.8cm

\begin{tabular}{|r|l|}
\hline
LO & \space \\
\hline
0.26 & Qinitial2 \\
\hline
0 & Order \\
\hline
0 & ThreeFlavorMode \\
\hline
1      &           AlphaDoSeparateThreshold  \\
\hline
1        &         DoBottomThreshold \\
\hline \hline
NLO & \space \\
\hline
0.40 &         Qinitial2  \\
\hline
1      &         Order  \\
\hline
0       &        ThreeFlavorMode  \\
\hline
1        &         AlphaDoSeparateThreshold    \\
\hline
1        &         AlphaUseExact   \\
\hline
1         &        DoBottomThreshold   \\
\hline \hline
NNLO & \space \\
\hline
0.40 &         Qinitial2  \\
\hline
2      &         Order  \\
\hline
0       &        ThreeFlavorMode  \\
\hline
1        &         AlphaDoSeparateThreshold    \\
\hline
1        &         AlphaUseExact   \\
\hline
1        &         NNLOmultiOrderCHARM   \\
\hline
1         &        DoBottomThreshold   \\
\hline
1         &        NLO4NNLO   \\
\hline
\end{tabular}
\vskip 0.8cm
In this mode it is necessary to generate LO and NLO sets 
by running the program before running the NNLO set on the same grid! 
Those will be dumped in special data files\\
({\bf agrv99lo.BO.threshold},
 {\bf agrv99lo.CH.threshold},\\
 {\bf agrv99nlo.BO.threshold},
and 
 {\bf agrv99nlo.CH.threshold})\\
that will later be read
for the NNLO calculation whenever \\
NNLOmultiOrderCHARM=1.

4. To generate VFNS densities involving proper order mixing at heavy thresholds
with all heavy flavors (both charm and bottom) and using LO, NLO and NNLO
(approximate) weights one sets:
\vskip 0.8cm

\begin{tabular}{|r|l|}
\hline
LO & \space \\
\hline
0.26 & Qinitial2 \\
\hline
0 & Order \\
\hline
0 & ThreeFlavorMode \\
\hline
1      &           AlphaDoSeparateThreshold  \\
\hline
1        &         DoBottomThreshold \\
\hline \hline
NLO & \space \\
\hline
0.40 &         Qinitial2  \\
\hline
1      &         Order  \\
\hline
0       &        ThreeFlavorMode  \\
\hline
1        &         AlphaDoSeparateThreshold    \\
\hline
1        &         AlphaUseExact   \\
\hline
1         &        DoBottomThreshold   \\
\hline \hline
NNLO & \space \\
\hline
0.40 &         Qinitial2  \\
\hline
2      &         Order  \\
\hline
0       &        ThreeFlavorMode  \\
\hline
1        &         AlphaDoSeparateThreshold    \\
\hline
1        &         AlphaUseExact   \\
\hline
1        &         NNLOmultiOrderCHARM   \\
\hline
1         &        DoBottomThreshold   \\
\hline
0         &        NLO4NNLO   \\
\hline
\end{tabular}
\vskip 0.8cm
In this mode it is also necessary to generate LO and NLO sets 
by running the program before running the NNLO set on the same grid! 
Those will be dumped in special data files\\
({\bf agrv99lo.BO.threshold},
 {\bf agrv99lo.CH.threshold},\\
 {\bf agrv99nlo.BO.threshold},
and
 {\bf agrv99nlo.CH.threshold})\\
that will later be read
for NNLO calculation whenever \\
NNLOmultiOrderCHARM=1.

Program output is arranged in several forms. First, the default output in
normal readable form goes into {\bf resLO.dat, resNLO.dat or resNNLO.dat} or
for GRV98-mode into {\bf resLO3.dat, resNLO3.dat or resNNLO3.dat } depending
upon the set calculation order. This file contains the input parameters,
calculation time and the columns of data versus $Q^2$ and $x$ for all densities
(uv,dv,us,ds,ss,ch and bt, described in previous chapters). Here is the
sample:

\begin{verbatim}

========================== Q2=   1.960 =======================
Alpha(Q2=    1.96 GeV2)=0.318513 for nf=4
-------------------------- x=0.000010  -----------------------
SI(x= 0.0000100)=3.4695646e+00  GL(x= 0.0000100)=1.3074834e+01
UV(x= 0.0000100)=6.1120367e-03  DV(x= 0.0000100)=3.7959190e-03
US(x= 0.0000100)=5.9818110e-01  DS(x= 0.0000100)=5.9988948e-01
SS(x= 0.0000100)=5.3175774e-01
CH(x= 0.0000100)=0.0000000e+00  BT(x= 0.0000100)=0.0000000e+00
-------------------------- x=0.000020  -----------------------
SI(x= 0.0000200)=3.1153438e+00  GL(x= 0.0000200)=1.1469641e+01
UV(x= 0.0000200)=8.2564168e-03  DV(x= 0.0000200)=5.1210226e-03
US(x= 0.0000200)=5.4149612e-01  DS(x= 0.0000200)=5.4372565e-01
SS(x= 0.0000200)=4.6576141e-01
CH(x= 0.0000200)=0.0000000e+00  BT(x= 0.0000200)=0.0000000e+00

\end{verbatim}
The above sample was produced with GraphAll=0 thus printing only values on a
small grid with minimum $Q^2=1.96$ ${\rm GeV}^2$ 
and not all values from minimum
$Q^2=0.40$ ${\rm GeV}^2$.
For convenience, SI denotes singlet, GL gluon, UV and DV 
are valence densities $u - \bar u$,  $d -\bar u$, US, DS, SS 
are of the $q + \bar q - \Sigma(n_f)/n_f$ kind
and CH and BT are $(c+\bar c)/2$ and $(b+\bar b)/2$.

For graphing purposes, the output also goes into several datafiles with 
names formed as {\bf g\_densityORDER.dat} where ORDER is LO, NLO or NNLO
respectively
e.g. {\bf g\_glLO.dat} or {\bf g\_uvNNLO.dat}.  Those contains
columns of the particular density with the first column being $x$ or $Q^2$,
depending upon GraphVsX parameter (1-$x$, 0-$Q^2$). Then the other parameter
is varied across columns. Here is the piece of {\bf g\_cpNLO.dat} file.
The first column contains the  $x$ value, the
second is the charm density for $Q^2=1.96$
${\rm GeV}^2$ (where it is zero) and then the charm density
for $Q^2=2,3,..$ ${\rm GeV}^2$:

\begin{verbatim}
0.0000100000 0.0000000000e+00   1.0468420825e-02   1.3022045486e-01
0.0000200000 0.0000000000e+00   8.8559755303e-03   1.0970003578e-01
0.0000300000 0.0000000000e+00   8.0049032415e-03   9.8909559249e-02   
0.0000400000 0.0000000000e+00   7.4395488912e-03   9.1758900075e-02   
0.0000500000 0.0000000000e+00   7.0219632243e-03   8.6487022224e-02   
0.0000600000 0.0000000000e+00   6.6940248888e-03   8.2352079221e-02   
0.0000700000 0.0000000000e+00   6.4257535493e-03   7.8973989377e-02   
0.0000800000 0.0000000000e+00   6.1998499386e-03   7.6132631291e-02   
0.0000900000 0.0000000000e+00   6.0054517691e-03   7.3690103826e-02   

\end{verbatim}
The above sample was produced with GraphVsX=1 thus printing $x$, 
not $Q^2$ values in the first column. 
The GraphAll=0 was also set, thus only nice values of $x$ are
used (0.00001, 0.00002, 0.00003, etc).

Also, if the necessary option (DoFortran=1) is set the output also goes 
into the file suitable for reading by a GRV98-like Fortran program 
that interpolates the data points and makes parton density functions. 
This program is used in structure function calculations 
(the code is written in Fortran). The datafile format has eight
columns with all densities on the fixed grid (hard-coded into the
both evolution code and the interpolation program) in $x$ and $Q^2$.

The sample follows:

\begin{verbatim}
Information line: first
+6.112E-03 +3.796E-03 +5.982E-01 +5.999E-01 +5.318E-01 +1.307E+01 
+6.128E-03 +3.806E-03 +6.084E-01 +6.101E-01 +5.419E-01 +1.333E+01 
+6.303E-03 +3.912E-03 +7.251E-01 +7.268E-01 +6.581E-01 +1.634E+01 
+6.440E-03 +3.996E-03 +8.260E-01 +8.278E-01 +7.586E-01 +1.901E+01 
+6.553E-03 +4.064E-03 +9.151E-01 +9.169E-01 +8.473E-01 +2.140E+01 
+6.649E-03 +4.122E-03 +9.949E-01 +9.967E-01 +9.269E-01 +2.357E+01 
+6.731E-03 +4.172E-03 +1.067E+00 +1.069E+00 +9.990E-01 +2.556E+01 
+6.804E-03 +4.216E-03 +1.134E+00 +1.135E+00 +1.065E+00 +2.740E+01 
+6.869E-03 +4.255E-03 +1.195E+00 +1.197E+00 +1.126E+00 +2.910E+01 
+6.927E-03 +4.291E-03 +1.252E+00 +1.253E+00 +1.183E+00 +3.070E+01  
\end{verbatim}

Sample pictures of bottom densities are provided in \cite{cbot} and also
below in Figs. 1 - 4.



%% file: my8.tex
\mysection{Error code descriptions}

Program error code description:
\vskip 0.8cm 

\begin{tabular}{|p{4cm}|l|p{5cm}|}
\hline                                    
message & filename & refer to \\
\hline
\hline                                    
Threshold LO datafile is missing      &   loader.c & NNLO calculation with
proper orders requires the datafile from a previous run in LO\\
\hline                                    
Threshold NLO datafile is missing     &  loader.c & NNLO calculation with
proper orders requires the datafile from a previous run in NLO \\
\hline                                    
Wrong Multicharm factor  &  main.c & NNLOmultiOrderCHARM should be 1 or 0\\
\hline                             
Wrong order factor &   several modules & should be 0,1,2 for LO, NLO, NNLO\\
\hline                             
File evolution\_parameters.input does not exist & main.c & find the file
and put into working directory \\
\hline                             
Wrong INI Q2:increase it!&main.c&order and initial $Q^2$ are incompatible \\
\hline                             
Wrong INI Q2:decrease it!&main.c&order and initial $Q^2$ are incompatible\\
\hline                             
Evolver: dont know how to proceed & main.c & wrong doBottom factor  \\
\hline                             
Wrong Alpha switch factor!&main.c & check AlphaDoSeparateThreshold value\\
\hline                             
Wrong order factor while graphing &  main.c & check Order to be 0,1,2 \\
\hline                             
Wrong graphing factor &   main.c &  check GraphAll value to be 0,1 \\
\hline                             
Wrong loadWgts factor & l-a-w.c, nl-a-w.c & check loadWgts to be 0,1  \\
\hline                             
\hline                             
\end{tabular}



%% file: my9.tex
\mysection{Conclusions}

We have presented a multifunctional code for the direct
$x$-space method of solving the spin-averaged  evolution
equations for parton densities. The distinctive
features of this code include analytic computation of the LO,
NLO and NNLO weights, NNLO heavy flavor
threshold matching  and NNLO evolution.

The code is very fast and accurate.
For example for grid sizes not exceeding 200 in $Q^2$ 
and 150 in $x$ the NLO calculation with full weighs computed for 
three values of $n_f$ and up to five decimal accuracy has 
a runtime well below 200 seconds. 
Also it is the only code that does the proper NNLO evolution with
NNLO heavy flavor matching conditions. 

The program is also easy to use and complete documentation is available.
The code is well-tested both on specific test functions 
(e.g. see \cite{brnv}) and on actual densities 
(e.g. see \cite{csn}) in all (LO, NLO and NNLO) orders.

\mysection{Acknowledgments}

The work  was partially supported
by the National Science Foundation grant PHY-9722101.
We thank Michael Botje for valuable comments on his method of solving the
evolution equations and providing us with his evolution code.
We also thank Andreas Vogt for help with testing the code in NNLO 
and for providing comparison data. Thanks are also due to
Brian Harris for testing the code and comments on the manuscript.


%% file: myaa.tex
\mysection{Appendix A}
\setcounter{section}{1}

Here we give the NNLO parametrizations of the splitting functions
from \cite{3loop2}.
Note that $L_0=\ln{z}$ and $L_1 = \ln(1-z)$. 

First the parametrizations for the non-singlet splitting 
functions $P^{(2)\pm} _{\rm NS}$ are:

\bea
  P^{(2)-}_{{\rm NS}, A}(z)\! &=&
   1185.229\:(1-z)_+^{-1} + 1365.458\: \delta (1-z) - 157.387\: L_1^2 
   - 2741.42\: z^2 
  \nonumber \\ & & \mbox{} 
   - 490.43\: (1-z) + 67.00\: L_0^2 + 10.005\: L_0^3 + 1.432\:  L_0^4
   \nonumber \\ & & \mbox{} \hspace{-13mm}
   + N_f\: \{ - 184.765\: (1-z)_+^{-1} - 184.289\: \delta (1-z)
   + 17.989\: L_1^2 + 355.636\: z^2 
  \nonumber \\ & & \mbox{} 
   - 73.407\: (1-z)L_1 + 11.491\: L_0^2 + 1.928\:  L_0^3 \} \:\: + \:\:
   P^{(2)}_{{\rm NS},2}(z) ,
  \nonumber \\
\label{A.1}
  P^{(2)-}_{{\rm NS}, B}(z)\! &=&
   1174.348\: (1-z)_+^{-1} + 1286.799\: \delta (1-z) + 115.099\: L_1^2 
   + 1581.05\: L_1  
  \nonumber \\ & & \mbox{}
   + 267.33\:  (1-z) - 127.65\: L_0^2 - 25.22\: L_0^3 + 1.432\: L_0^4
  \nonumber \\ & & \mbox{} \hspace{-13mm}
   + N_f\: \{
   - 183.718\: (1-z)_+^{-1} - 177.762\: \delta (1-z)
   + 11.999\: L_1^2 + 397.546\: z^2 
  \nonumber \\ & & \mbox{} 
   + 41.949\: (1-z) - 1.477\: L_0^2 - 0.538\:  L_0^3 \} \:\: + \:\:
  P^{(2)}_{{\rm NS},2}(z)  ,
\eea
and
\bea
  P^{(2)+}_{{\rm NS}, A}(z)\! &=&
   1183.762\: (1-z)_+^{-1} + 1347.032\: \delta (1-z) + 1047.590\: L_1 
   - 843.884\: z^2
  \nonumber \\ & & \mbox{}
   - 98.65\: (1-z) - 33.71\: L_0^2 + 1.580\: (L_0^4 + 4L_0^3)
  \nonumber \\ & & \mbox{} \hspace{-13mm}
   + N_f\: \{ - 183.148\: (1-z)_+^{-1} - 174.402\: \delta (1-z)
   + 9.649\: L_1^2 + 406.171\: z^2 
  \nonumber \\ & & \mbox{} 
   + 32.218\: (1-z) + 5.976\: L_0^2 + 1.60\:  L_0^3 \} \:\: + \:\:
   P^{(2)}_{{\rm NS},2}(z)  ,
  \nonumber \\
\label{A.2}
  P^{(2)+}_{{\rm NS}, B}(z)\! &=&
   1182.774\: (1-z)_+^{-1} + 1351.088\: \delta (1-z) - 147.692\: L_1^2 
   - 2602.738\: z^2   
  \nonumber \\ & & \mbox{}
   - 170.11 + 148.47\: L_0 + 1.580\: (L_0^4 - 4\, L_0^3)
  \nonumber \\ & & \mbox{} \hspace{-13mm}
   + N_f\: \{ - 183.931\: (1-z)_+^{-1} - 178.208\: \delta (1-z)
   - 89.941\: L_1 + 218.482\: z^2 
  \nonumber \\ & & \mbox{} 
   + 9.623  + 0.910\: L_0^2 - 1.60\:  L_0^3 \} \:\: + \:\:
  P^{(2)}_{{\rm NS},2}(z) \:\: .
\eea
The parametrizations for $P_{\rm NS}^{(2),S}(z)$ and $P_{\rm PS}^{(2)}(z)$ are
\bea
  P^{(2)S}_{{\rm NS}, A}(z)\! &=&
   N_f\: \{ (1-z) ( - 1441.57\: z^2 + 12603.59\: z - 15450.01) 
   + 7876.93\: zL_0^2 
  \nonumber \\ & & \mbox{} 
   - 4260.29\: L_0 - 229.27\: L_0^2 + 4.4075\: L_0^3 \}
  \nonumber\\
\label{A.3}
  P^{(2)S}_{{\rm NS}, B}(z)\! &=&
   N_f\: \{ (1-z) ( -704.67 \: z^3 + 3310.32\: z^2 + 2144.81\: z 
   - 244.68) 
  \nonumber \\ & & \mbox{} 
   + 4490.81\: z^2 L_0 + 42.875\: L_0 - 11.0165\: L_0^3 \}  ,
\eea
and
\bea
  P^{(2)}_{{\rm PS}, A}(z)\! &=&
   N_f\: \{ (1-z)(-229.497\: L_1 - 722.99\: z^2 + 2678.77 
   - 560.20\: z^{-1})
  \nonumber \\ & & \mbox{}
   + 2008.61\: L_0 + 998.15\: L_0^2 - 3584/27\: z^{-1}L_0 \}
   \:\: + \:\: P^{(2)}_{{\rm PS},2}(z)  ,
  \nonumber\\
\label{A.4}
  P^{(2)}_{{\rm PS}, B}(z)\! &=&
   N_f\: \{ (1-z) (73.845\: L_1^2 + 305.988\: L_1 + 2063.19\: z 
   - 387.95\: z^{-1}) 
  \nonumber \\ & & \mbox{}
   + 1999.35\: zL_0 - 732.68\: L_0 - 3584/27\: z^{-1}L_0 \}
  \nonumber \\ & & \mbox{}
   \:\: + \:\: P^{(2)}_{{\rm PS},2}(z)  ,
\eea
with
\bea
\label{A.5}
  P^{(2)}_{{\rm PS},2}(z)\! & = & 
   N_f^2\: \{ (1-z)(-7.282\: L_1 - 38.779\: z^2 + 32.022\: z - 6.252
   + 1.767\: z^{-1} ) \quad
  \nonumber \\ & & \mbox{}
   + 7.453\: L_0^2 \} \:\: .
\eea

Next we show the parametrizations of the 
off-diagonal singlet splitting functions:

\bea
  P^{(2)}_{qg, A}(z)\! &=&
   N_f\: \{ - 31.830\: L_1^3 + 1252.267\: L_1 + 1999.89\: z + 1722.47  
   + 1223.43\: L_0^2
  \nonumber \\ & & \mbox{} 
    - 1334.61\: z^{-1} - 896/3\: z^{-1}L_0 \}
   \:\: + \:\: P^{(2)}_{qg,2}(z)  ,
  \nonumber\\
\label{A.6}
  P^{(2)}_{qg, B}(z)\! &=&
   N_f\: \{ 19.428\: L_1^4 + 159.833\: L_1^3 + 309.384\: L_1^2 
   + 2631.00\: (1-z) 
  \nonumber \\ & & \mbox{} 
   - 67.25\: L_0^2 - 776.793\: z^{-1} - 896/3\: z^{-1}L_0 \}
   \:\: + \:\: P^{(2)}_{qg,2}(z)  ,
\eea
with
\bea
\label{A.7}
  P^{(2)}_{qg,2}(z)\! &=&
   N_f^2\: \{ - 0.9085\: L_1^2 - 35.803\: L_1 - 128.023 + 200.929\: 
   (1-z) 
  \nonumber \\ & & \mbox{} 
   + 40.542\: L_0 + 3.284\: z^{-1} \} \:\: ,
\eea
and 
\bea
  P^{(2)}_{gq, A}(z)\! &=&
   13.1212\: L_1^4 + 126.665\: L_1^3 + 308.536\: L_1^2 + 361.21
   - 2113.45\: L_0 
  \nonumber \\ & & \mbox{} - 17.965\: z^{-1}L_0
   \:\: + \:\: N_f\: \{
   2.4427\: L_1^4 + 27.763\: L_1^3 + 80.548\: L_1^2 
  \nonumber \\ & & \mbox{} 
   - 227.135 - 151.04\: L_0^2 + 65.91\: z^{-1}L_0 \} \:\: + \:\:
   P^{(2)}_{gq ,2}(z)  ,
  \nonumber\\
\label{A.8}
  P^{(2)}_{gq, B}(z)\! &=& \,
   - 4.5108\: L_1^4 - 66.618\: L_1^3 - 231.535\: L_1^2 - 1224.22\:
   (1-z) + 240.08\: L_0^2 \quad
  \nonumber \\ & & \mbox{}
   + 379.60\: z^{-1} (L_0 +4) \: + \:N_f \{
   - 1.4028\: L_1^4 - 11.638\: L_1^3 + 164.963\: L_1 
  \nonumber \\ & & \mbox{} 
   - 1066.78\: (1-z)- 182.08\: L_0^2 + 138.54\: z^{-1} (L_0 +2) \} 
  \nonumber \\ & & \mbox{} 
   \:\: + \:\: P^{(2)}_{gq ,2}(z)  ,
\eea
with
\bea
\label{A.9}
  P^{(2)}_{gq, 2}(z)\! & = &
   N_f^2\: \{ 1.9361\: L_1^2 + 11.178\: L_1 + 11.632 - 15.145\: (1-z) 
   + 3.354\: L_0 
  \nonumber \\ & & \mbox{} 
   - 2.133\: z^{-1} \} \:\: .
\eea

Last we show the parametrizations of the diagonal singlet splitting functions

\bea
  P^{(2)}_{gg, A}(z)\! &=&
   2626.38\: (1-z)_+^{-1} + 4424.168\: \delta (1-z) - 732.715\: L_1^2 
   - 20640.069\: z
  \nonumber \\ & & \mbox{}
   - 15428.58\: (1-z^2) - 15213.60\: L_0^2 + 16700.88\: z^{-1} 
   + 2675.85\: z^{-1}L_0
  \nonumber \\ & & \mbox{} \hspace{-13mm}
   + N_f\: \{ - 415.71\: (1-z)_+^{-1} - 548.569\: \delta (1-z)
   - 425.708\: L_1 + 914.548\: z^2 
  \nonumber \\ & & \mbox{} 
   - 1122.86 - 444.21\: L_0^2 + 376.98\: z^{-1} 
   + 157.18\: z^{-1}L_0 \} \:\:
  \nonumber \\ & & \mbox{} 
   + \:\: P^{(2)}_{gg,2}(z)  ,
  \nonumber\\
\label{A.10}
  P^{(2)}_{gg, B}(z)\! &=&
   2678.22\: (1-z)_+^{-1} + 4590.570\: \delta (1-z)
   + 3748.934\: L_1 + 60879.62\: z
  \nonumber \\ & & \mbox{}
   - 35974.45\: (1+z^2) + 2002.96\: L_0^2 + 9762.09\: z^{-1}
   + 2675.85\: z^{-1}L_0
  \nonumber \\ & & \mbox{} \hspace{-13mm}
   + N_f\: \{
   - 412.00\: (1-z)_+^{-1} - 534.951\: \delta (1-z)
   + 62.630\: L_1^2 + 801.90 
  \nonumber \\ & & \mbox{} 
   + 1891.40\: L_0 + 813.78\: L_0^2 + 1.360\: z^{-1} 
   + 157.18\: z^{-1}L_0 \} 
  \nonumber \\ & & \mbox{} 
   + \:\: P^{(2)}_{gg ,2}(z)  ,
   \quad
\eea
with
\bea
\label{A.11}
  P^{(2)}_{{gg}, 2}(z)\! & = &
   N_f^2\: \{ -16/9\: (1-z)_+^{-1} + 6.4882\: \delta (1-z)
   + 37.6417\: z^2 - 72.926\: z 
  \nonumber \\ & & \mbox{} 
   + 32.349 - 0.991\: L_0^2 + 2.818\: z^{-1} \} \:\: .
\eea
%

%% file: mybb.tex
\mysection{Appendix B}
\setcounter{section}{2}

Shown below are the renormalized OME's used for threshold matching 
calculations in NLO and NNLO (they correspond to the 
unrenormalized expressions 
given in Appendix C of \cite{bmsmn} and in Appendix A of \cite{bmsn1}).
All OME'S have been renormalized in the ${\overline {\rm MS}}$-scheme.

In particular the renormalized coupling $\alpha_s$ is presented in the
above scheme for $n_f+1$ light flavors. 
Here the heavy quark $H=(c,b)$ is treated
on the same footing as the light flavors and it is not decoupled
from the running coupling in the VFNS approach. 
The $(\alpha_s/4\pi)^2$ coefficient in the heavy-quark OME 
$\tilde A^{\rm PS}_{Hq}$ is given by

\begin{eqnarray}
&& \hskip-0.5cm \tilde A^{{\rm PS},(2)}_{Hq}\Biggl(\frac{m^2}{\mu^2}\Biggr)=
C_FT_f\Biggl\{
\Biggl[-8(1+z)\ln z-\frac{16}{3z}-4
\nonumber \\ && \qquad
+ 4 z +\frac{16}{3}z^2\Biggr] \ln^2\frac{m^2}{\mu^2}
+\Biggl[8(1+z)\ln^2z-\Biggl(8+40z+\frac{64}{3}z^2\Biggr)\ln z
\nonumber \\ && \qquad
-\frac{160}{9z}
+16-48z+\frac{448}{9}z^2\Biggr] \ln\frac{m^2}{\mu^2}
\nonumber \\ && \qquad
+ (1+z)\Biggl[32{\rm S}_{1,2}(1-z)+16\ln z{\rm Li}_2(1-z)
-16\zeta(2)\ln z
\nonumber \\ && \qquad
-\frac{4}{3}\ln^3z\Biggr]
+\Biggl(\frac{32}{3z}+8-8z-\frac{32}{3}z^2\Biggr) {\rm Li}_2(1-z)
\nonumber \\ && \qquad
+ \Biggl( -\frac{32}{3 z}-8+8z+\frac{32}{3} z^2\Biggr)\zeta(2)
+\Biggl(2+10z+\frac{16}{3}z^2\Biggr)  \ln^2z
\nonumber \\ && \qquad
-\Biggl(\frac{56}{3}+\frac{88}{3}z
+\frac{448}{9}z^2\Biggr)\ln z
-\frac{448}{27z} - \frac{4}{3}
-\frac{124}{3}z+\frac{1600}{27}z^2 \Biggr\}  \,,
\end{eqnarray}

The $\alpha_s/4\pi$ and the $(\alpha_s/4\pi)^2$ coefficients of the
heavy quark OME's $\tilde A^{\rm S}_{Hg}$ are

\begin{eqnarray}
&& 
\tilde A_{Hg}^{{\rm S},(1)} \Biggl(\frac{m^2}{\mu^2}\Biggr)
= T_f \Biggl[ - 4 ( z^2 + (1 - z)^2)\ln\frac{m^2}{\mu^2}\Biggr] \, ,
\end{eqnarray}
and

\begin{eqnarray}
&& \tilde A_{Hg}^{{\rm S},(2)}\Biggl(\frac{m^2}{\mu^2}\Biggr)
=
\Biggl\{C_FT_f[ (8 -16 z+16 z^2)\ln(1-z)
\nonumber \\ && \qquad
-(4 -8 z+ 16 z^2)\ln z -(2 - 8 z)]
\nonumber \\ && \qquad
+C_AT_f\Biggl[-(8 - 16 z + 16 z^2)\ln(1-z)
-(8 + 32 z)\ln z
\nonumber \\ && \qquad
 -\frac{16}{3z} -4  - 32 z+\frac{124}{3}z^2\Biggr]
+ T_f^2 \Biggl[ - \frac{16}{3} ( z^2 + (1 - z)^2) \Biggr]
\Biggr\} \ln^2\frac{m^2}{\mu^2}
\nonumber \\ && \qquad
+\Biggl\{C_FT_f \Biggl[( 8 - 16 z + 16z^2)[2\ln z\ln(1-z)
-\ln^2(1-z)+2\zeta(2)]
\nonumber \\ && \qquad
-(4 - 8 z +16 z^2)\ln^2z-32z(1-z)\ln(1-z)
\nonumber \\ && \qquad
-(12 - 16 z + 32 z^2)\ln z  - 56+116z -80z^2 \Biggr]
\nonumber \\ && \qquad
+ C_AT_f\Biggl[(16 +32 z +32 z^2)[{\rm Li}_2(-z) + \ln z\ln(1+z) ]
\nonumber \\ && \qquad
+(8 - 16 z + 16 z^2)\ln^2(1-z)
+(8 + 16 z)\ln^2z
\nonumber \\ && \qquad
+32z\zeta(2)+32z(1-z)\ln(1-z)
-\Biggl(8+64z+\frac{352}{3}z^2\Biggr)\ln z
\nonumber \\ && \qquad
-\frac{160}{9z}+16-200z+\frac{1744}{9}z^2 \Biggr]\Biggl\} \ln \frac{m^2}{\mu^2}
\nonumber \\ && \qquad
+ C_FT_f\Big\{(1-2z+2z^2)[8\zeta(3)
+\frac{4}{3}\ln^3(1-z)
\nonumber \\ && \qquad
-8\ln(1-z){\rm Li}_2(1-z)
+8\zeta(2)\ln z
-4\ln z\ln^2(1-z)
\nonumber \\ && \qquad
+\frac{2}{3}\ln^3z
-8\ln z{\rm Li}_2(1-z)
+8{\rm Li}_3(1-z)
-24{\rm S}_{1,2}(1-z)]
\nonumber \\ && \qquad
+z^2\Biggl[-16\zeta(2)\ln z+\frac{4}{3}\ln^3z
+16\ln z{\rm Li}_2(1-z)+32{\rm S}_{1,2}(1-z)\Biggr]
\nonumber \\ && \qquad
-(4+96z-64z^2){\rm Li}_2(1-z)
-(4-48z+40z^2)\zeta(2)
\nonumber \\ && \qquad
-(8+48z-24z^2)\ln z\ln(1-z)
+(4+8z-12z^2)\ln^2(1-z)
\nonumber \\ && \qquad
-(1+12z-20z^2)\ln^2z-(52z-48z^2)\ln(1-z)
\nonumber \\ && \qquad
-(16+18z+48z^2)\ln z
+26-82z+80z^2\Biggr\}
\nonumber \\ && \qquad
+C_AT_f\Biggl\{(1-2z+2z^2) [
-\frac{4}{3} \ln^3(1-z)
\nonumber \\ && \qquad
+8\ln(1-z){\rm Li}_2(1-z)-8{\rm Li}_3(1-z)]
+(1+2z+2z^2)
\nonumber \\ && \qquad
\times [-8\zeta(2)\ln(1+z)
-16\ln(1+z){\rm Li}_2(-z)
-8\ln z\ln^2(1+z)
\nonumber \\ && \qquad
+4\ln^2z\ln(1+z) + 8\ln z{\rm Li}_2(-z)-8{\rm Li}_3(-z)
-16{\rm S}_{1,2}(-z)]
\nonumber \\ && \qquad
+(16+64z)[2{\rm S}_{1,2}(1-z)
+\ln z{\rm Li}_2(1-z)]
-\Biggl(\frac{4}{3} +  \frac{8}{3} z\Biggr)\ln^3z
\nonumber \\ && \qquad
+(8-32z+16z^2)\zeta(3)-(16+64z)\zeta(2)\ln z+(16+16z^2)
\nonumber \\ && \qquad
\times [ {\rm Li}_2(-z) + \ln z\ln(1+z)  ]
+\Biggl(\frac{32}{3z}+12+64z-\frac{272}{3}z^2\Biggr)
{\rm Li}_2(1-z)
\nonumber \\ && \qquad
-\Biggl( 12 + 48 z - \frac{260}{3} z^2+\frac{32}{3 z}\Biggr)\zeta(2)
-4z^2\ln z\ln(1-z)
\nonumber \\ && \qquad
-(2+8z-10z^2)\ln^2(1-z)+\Biggl(2+8z+\frac{46}{3}z^2\Biggr)\ln^2z
\nonumber \\ && \qquad
+(4+16z-16z^2)\ln(1-z)
-\Biggl(\frac{56}{3}+\frac{172}{3}z+\frac{1600}{9}z^2\Biggr)\ln z
\nonumber \\ && \qquad
-\frac{448}{27z}-\frac{4}{3}-\frac{628}{3}z
+\frac{6352}{27}z^2\Biggr\} \, ,
\end{eqnarray}
respectively.
Now we present the renormalized expressions for 
the heavy-quark loop contributions to the light-parton OME's denoted by
$A_{kl,H}$. The coefficients of the $(\alpha_s/4\pi)^2$ terms in 
$A_{qq,H}$ and $A_{gq,H}$ are
\begin{eqnarray}
&&  A^{{\rm NS},(2)}_{qq,H}\Biggl(\frac{m^2}{\mu^2}\Biggr)
=
 C_F T_f \Biggl\{\Biggr[
\frac{8}{3}\Biggl(\frac{1}{1-z}\Biggr)_+
-\frac{4}{3}-\frac{4}{3}z+2\delta(1-z)\Biggr] \ln^2\frac{m^2}{\mu^2}
\nonumber \\ && \qquad
+\Biggl[\frac{80}{9}\Biggl(\frac{1}{1-z}\Biggr)_+ +\frac{8}{3}
\frac{1+z^2}{1-z}\ln z+\frac{8}{9}-\frac{88}{9}z
\nonumber \\ && \qquad
+\delta(1-z)\Biggl(
\frac{16}{3}\zeta(2)+\frac{2}{3}\Biggr)\Biggr] \ln\frac{m^2}{\mu^2}
\nonumber \\ && \qquad
+\frac{1+z^2}{1-z}\Biggl(\frac{2}{3}\ln^2z+\frac{20}{9}\ln z\Biggr)
\nonumber \\ && \qquad
+\frac{8}{3}(1-z)\ln z
+\frac{224}{27}\Biggl(\frac{1}{1-z}\Biggr)_+
+\frac{44}{27}-\frac{268}{27}z
\nonumber \\ && \qquad
+\delta(1-z)\Biggl(-\frac{8}{3}\zeta(3)+\frac{40}{9}
\zeta(2)+\frac{73}{18}\Biggr) \Biggr\} \,,
\end{eqnarray}
and  
\begin{eqnarray}
&&A_{gq,H}^{{\rm S},(2)}\Biggl(\frac{m^2}{\mu^2}\Biggr) =
C_FT_f \Biggl\{ \Biggl[
\frac{16}{3 z} - \frac{16}{3} + \frac{8}{3} z \Biggr] \ln^2 \frac{m^2}{\mu^2}
\nonumber \\ && \qquad
+\Biggl[\frac{160}{9 z} - \frac{160}{9} + \frac{128}{9} z
+ (\frac{32}{3 z} - \frac{32}{3} + \frac{16}{3} z) \ln(1 - z) \Biggr]
\ln\frac{m^2}{\mu^2}
\nonumber \\ && \qquad
+ \frac{4}{3} \Biggl(\frac{2}{z} - 2 + z \Biggr) \ln^2(1 - z)
+ \frac{8}{9}  \Biggl(\frac{10}{z} - 10 + 8 z \Biggr) \ln(1 - z)
\nonumber \\ && \qquad
+ \frac{1}{27} \Biggl(\frac{448}{z} - 448 + 344 z \Biggr )
\Biggr\} \, .
\end{eqnarray}
respectively. The coefficients of the $\alpha_s/4\pi$ and
$(\alpha_s/4\pi)^2$ terms in $A_{gg,H}$ are 
\begin{eqnarray}
&&   
A_{gg,H}^{{\rm S},(1)} \Biggl(\frac{m^2}{\mu^2}\Biggr)
=T_f\Biggl[\frac{4}{3}\delta(1-z) 
\ln\frac{m^2}{\mu^2} \Biggr] \, ,
\end{eqnarray}                      
and
\begin{eqnarray}
&&A_{gg,H}^{{\rm S},(2)}\Biggl(\frac{m^2}{\mu^2}\Biggr) =
\Biggl\{C_FT_f\Biggl[8 (1+z)\ln z
+\frac{16}{3 z} + 4
- 4 z - \frac{16}{3} z^2  \Biggr]
\nonumber \\ && \qquad
+C_AT_f \Biggl[
\frac{8}{3}\Biggl(\frac{1}{1-z}\Biggr)_+
+ \frac{8}{3 z} - \frac{16}{3} + \frac{8}{3} z - \frac{8}{3} z^2 \Biggr]
\nonumber \\ && \qquad
+T_f^2 \Biggl[\frac{16}{9}\delta(1-z) \Biggr]
\Biggr\} \ln^2 \frac{m^2}{\mu^2}
\nonumber \\ && \qquad
+\Biggl\{C_FT_f \Biggl[
8 (1 + z) \ln^2 z + (24 + 40 z) \ln z - \frac{16}{3 z} + 64
- 32 z
\nonumber \\ && \qquad
- \frac{80}{3} z^2  + 4 \delta(1-z)\Biggr]
+ C_AT_f\Biggl[
\frac{16}{3} (1 + z) \ln z + \frac{80}{9}\Biggl(\frac{1}{1-z}\Biggr)_+
\nonumber \\ && \qquad
+ \frac{184}{9 z} - \frac{232}{9} + \frac{152}{9} z
- \frac{184}{9} z^2 + \frac{16}{3} \delta(1- z)
 \Biggr]\Biggr\} \ln \frac{m^2}{\mu^2}
\nonumber \\ && \qquad
+ C_F T_f\Biggl\{
\frac{4}{3}(1 + z)\ln^3 z
+(6 + 10 z) \ln^2 z + (32 + 48 z) \ln z
\nonumber \\ && \qquad
\nonumber \\ && \qquad
- \frac{8}{z} + 80 - 48 z - 24 z^2 - 15 \delta(1 - z)
\Biggr\}
\nonumber \\ && \qquad
+C_AT_f\Biggl\{\frac{4}{3}(1 + z)\ln^2 z
+ \frac{1}{9} (52 + 88 z) \ln z
- \frac{4}{3} z \ln(1-z)
\nonumber \\ && \qquad
+\frac{1}{27} \Biggl[224 \Biggl(\frac{1}{1-z}\Biggr)_+ + \frac{556}{z}
- 628 + 548 z - 700 z^2 \Biggr]
\nonumber \\ && \qquad
+ \frac{10}{9} \delta(1 - z)
\Biggr\} \, ,
\end{eqnarray}
respectively.

The definitions for the polylogarithms ${\rm Li}_n(z)$ and the
Nielsen functions ${\rm S}_{n,p}(z)$, which appear in the
above expressions, can be found in \cite{lbmr}.


%% file: myr.tex
%

%% file: myf.tex
\centerline{\bf \large{Figure Captions}}
\begin{description}
\item[Fig. 1.]
The gluon density $xg_{\rm NNLO}(4,x,\mu^2)$ in the range
$10^{-5} < x < 1$ for 
$\mu^2 =$ 2, 3, 4, 5, 10 and 20 in units 
of $({\rm GeV}^2)^2$, 
\item[Fig. 2.]
The singlet density $x\Sigma_{\rm NNLO}(4,x,\mu^2)$ 
in the range $10^{-5} < x < 1$ for 
$\mu^2 =$ 2, 3, 4, 5, 10 and 20 in units 
of $({\rm GeV}^2)^2$, 
\item[Fig. 3.]
The nonsinglet quark density $x\sigma_{\rm NNLO}(4,x,\mu^2)a,$  
where $\sigma = (u + \bar u)/2$,
in the range $10^{-5} < x < 1$ for 
$\mu^2 =$ 2, 3, 4, 5, 10 and 20 in units 
of $({\rm GeV}^2)^2$, 
\item[Fig. 4.]
The charm quark density $xc_{\rm NNLO}(4,x,\mu^2)$ the range
$10^{-5} < x < 1$ for 
$\mu^2 =$ 1.96, 2, 3, 4, 5, 10 and 20 in units 
of $({\rm GeV}^2)^2$, 
\end{description}